\documentclass[preprint,aps]{revtex4}
\usepackage{graphicx}
\usepackage{amsmath}

\newcommand{\beq}{\begin{equation}}
\newcommand{\enq}{\end{equation}}
\newcommand{\bea}{\begin{eqnarray}}
\newcommand{\ena}{\end{eqnarray}}
\newcommand{\rr}{{\bf r}}
\newcommand{\aos}{a_{\rm osc}}
\newcommand{\pmin}{p_{\rm min}}
\newcommand{\pmax}{p_{\rm max}}

\def\vec#1{{\mathbf #1}}
\def\hat#1{{#1}^{*}}

\def\conjg#1{{#1}^{*}}

\begin{document}
\title{Long term dynamics of the splitting of a doubly quantized vortex in
  a two-dimensional condensate}
\author{Halvor M.\ Nilsen}
\affiliation{Centre of Mathematics for Applications, 
P.O. Box 1053 Blindern, NO-0316 Oslo, Norway}
\author{Emil Lundh}
\affiliation{Centre of Mathematics for Applications, 
P.O. Box 1053 Blindern, NO-0316 Oslo, Norway}
\affiliation{Department of Physics, Ume{\aa} University, 
SE-90187 Ume{\aa}, Sweden\footnote{Present address.}}

\begin{abstract}
We study the nonlinear dynamics of the splitting of a doubly quantized
vortex in a trapped condensate. The dynamics is studied in 
detail by solving the 
Gross-Pitaevskii equation. The main 
dynamical features are explained in terms of a 
nonlinear
three-level 
system. We find an analytical solution for the characteristics of the
dynamics.
It is concluded that the time scale for the splitting is mainly
determined by the instability of the linearized system, and nonlinear 
effects contribute logarithmically. 
\end{abstract}

\maketitle

\section{Introduction}
Quantization of fluid circulation is one of the most pictorial 
macroscopic manifestations of quantum mechanics. Lattices of 
singly quantized vortices have been imaged in superconductors 
in magnetic fields \cite{essmann}, liquid helium \cite{vinen}, 
and more recently in trapped Bose-Einstein condensates 
\cite{madison2000,ketterle2001}. Vortices with higher quantum numbers 
than unity are energetically unstable in many common situations, 
including an infinite, homogeneous s-wave superfluid, and 
the experimentally relevant case of a condensate 
contained in a parabolic potential \cite{Pethick2001,butts1999,lundh2002}. 
In addition, in the latter case, multiply quantized vortices are 
found to be {\em dynamically} unstable in large areas of parameter 
space \cite{pu1999,jackson2005,mottonen2003}. It was predicted that 
a doubly quantized vortex is unstable towards splitting into 
two vortices with unit quantum number, in accordance with the 
quantization of fluid circulation.

These predictions were put to an experimental test in 2004, when a 
doubly quantized vortex, i.e., a  
vortex with quantum number 2, was imposed on a stationary condensate 
and the subsequent splitting was monitored \cite{shin2004}. This 
experiment has been analyzed quantitatively in Refs.\ 
\cite{mottonen2006,mateo2006} using the time-dependent Gross-Pitaevskii 
equation \cite{Pethick2001}, and in Refs.\ \cite{Lundh06,huhtamaki2006} by 
means of Bogoliubov theory. However, there remains to marry together 
these two approaches. In particular, Bogoliubov analysis gives 
information only about the linear (i.~e., short-time) behavior of 
the unstable system, while solving the full Gross-Pitaevskii 
equation gives more detail than is necessary in order to understand 
the important features of the dynamics.

Dynamics of vortices is a subject with a long history. It is well known
that in a incompressible fluid the vortices move with the background
fluid velocity \cite{kelvin}. This is not so in a compressible fluid
where the background density changes \cite{Nilsen06}. In general, 
vortex motion 
in a compressible fluid 
is complicated 
and cannot be
separated from the dynamics of the system. The splitting of
a doubly quantized vortex offers an opportunity to study the vortex 
dynamics in an extreme regime where
the background velocity changes rapidly on the scale of the 
size of a vortex
core. The splitting dynamics therefore offers insight into 
compressible fluid dynamics. In the study of the 
linear stability of doubly quantized vortices \cite{Lundh06}, it was
shown that the stability depends critically on the energy of the
surface modes, and thus on global properties not associated with the
vortex. The focus of this paper will be on the dynamics after the
initial exponential growth of the vortex distance. Even though the 
experiment of Ref.\ \cite{shin2004} was performed in an elongated 
three-dimensional geometry, this study is concerned with a 
two-dimensional system, in order to clearly bring out the 
structure of the problem.

In this paper, we perform a systematic investigation of the long time
behavior of the splitting of two vortices. 
The paper is organized as follows. In Sec.\
\ref{sec:genneral} we discuss the equations governing the system.
In Sec.\ \ref{sec:numerical_method} we describe the numerical 
solution of the equations of motion. 
Section \ref{sec:nummerical_calculations} is devoted to a
calculation of the nonlinear dynamics. The main 
features of the dynamics are captured in terms of a model that is solved
analytically in Sec.\ \ref{sec:vortex_dynamic_model}. 
Finally, in Sec.\ \ref{sec:conclusion} we summarize and
conclude. Specifics of the analytical solution are given in the three 
appendices.

\section{Splitting of a doubly quantized vortex} 
\label{sec:genneral}
The system we study is a Bose-Einstein condensate of particles of 
mass $m$ that is trapped in a
cylindrically symmetric potential. At zero temperature in the dilute
limit the gas is described by a condensate wavefunction 
$\Psi(\vec{r},t)$ that obeys the Gross-Pitaevskii (GP) equation 
\begin{equation}
i\hbar \frac{\partial \Psi}{\partial t} =
H_0\Psi +
U_0|\Psi|^2\Psi, 
\label{eqn:GPE}
\end{equation}
where 
\beq
H_0 = -\frac{\hbar^2}{2m}\nabla^2+ V(\rr),
\enq
and the trapping potential is assumed to be of the form
\begin{equation}
V({\rr}) = \frac{m\omega^2}{2}(r^2+\lambda^2 z^2).
\end{equation}
The inter-particle interactions are parametrized by an $s$-wave scattering 
length $a$, so that $U_0=4\pi\hbar^2a/m$. 
We immediately pass to trap units, where the unit of length is the 
oscillator length $\aos=(\hbar/m\omega)^{1/2}$ and the unit of time 
is $\omega^{-1}$ \cite{Pethick2001}. 
We assume the system to be 
two-dimensional (2D), which corresponds to the limit of a very tight trapping 
potential in the axial direction. The wavefunction in that direction is 
thus assumed to be in the ground state; on integrating out the $z$ 
dependence one obtains the effective 2D interaction parameter 
\beq
C =\frac{Na}{\aos}\int |\phi_0(z)|^4 {\rm d} z = 
\frac{Na\sqrt{\lambda}}{\aos\sqrt{2 \pi}}, 
\label{dimensionless}
\enq
where $\phi_0$ is the ground-state single-particle wave function in 
a one-dimensional harmonic potential. 
The resulting 
equation of motion for the condensate is 
\begin{equation}
\label{eq:dimlessgpe}
i\frac{\partial \Psi}{\partial t} =
\left[-\frac{1}{2}\nabla^2+ \frac12 r^2
+C|\Psi|^2\right]\Psi. 
\end{equation}

\begin{figure}[ht]
\includegraphics[width=0.4\columnwidth]{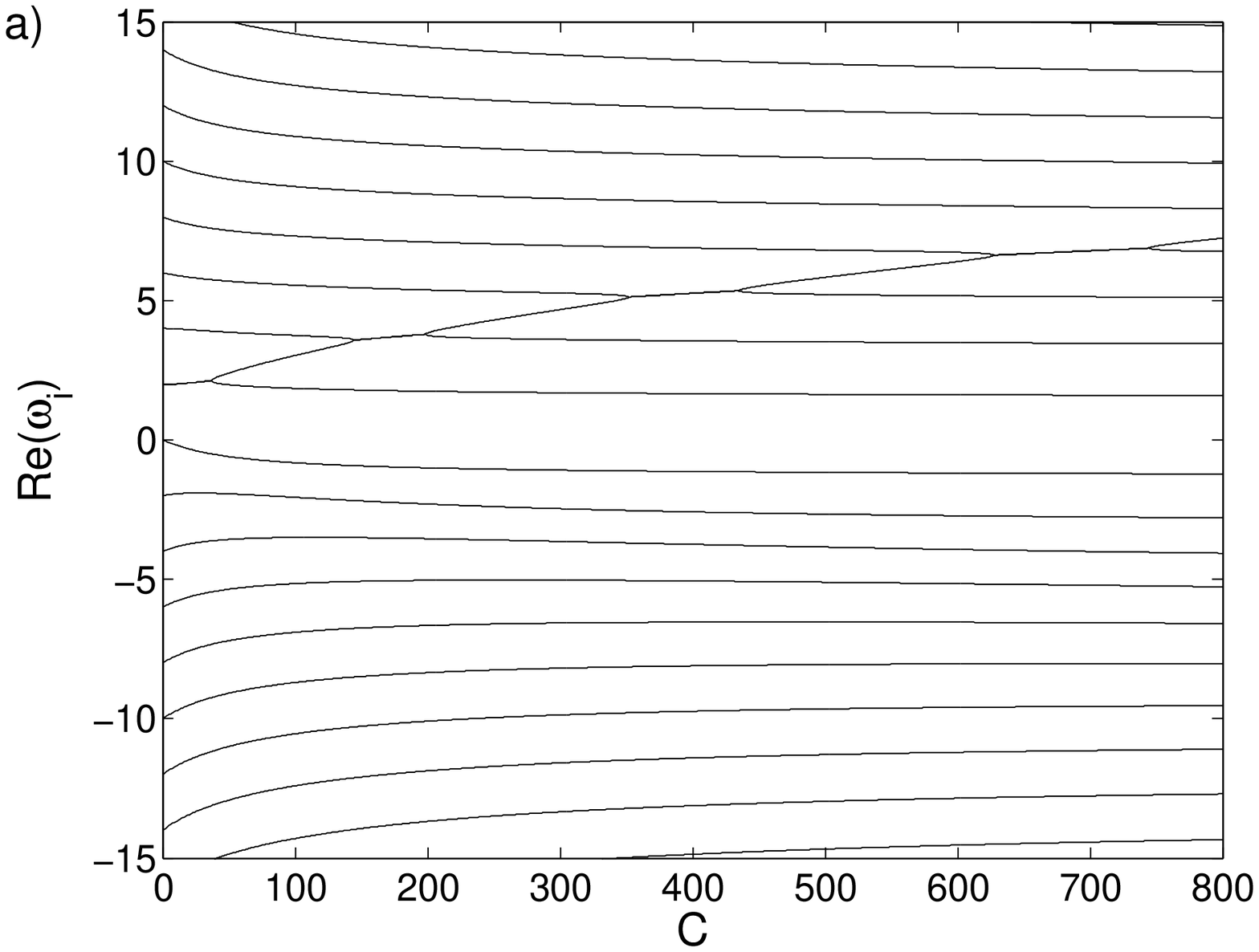}
\includegraphics[width=0.4\columnwidth]{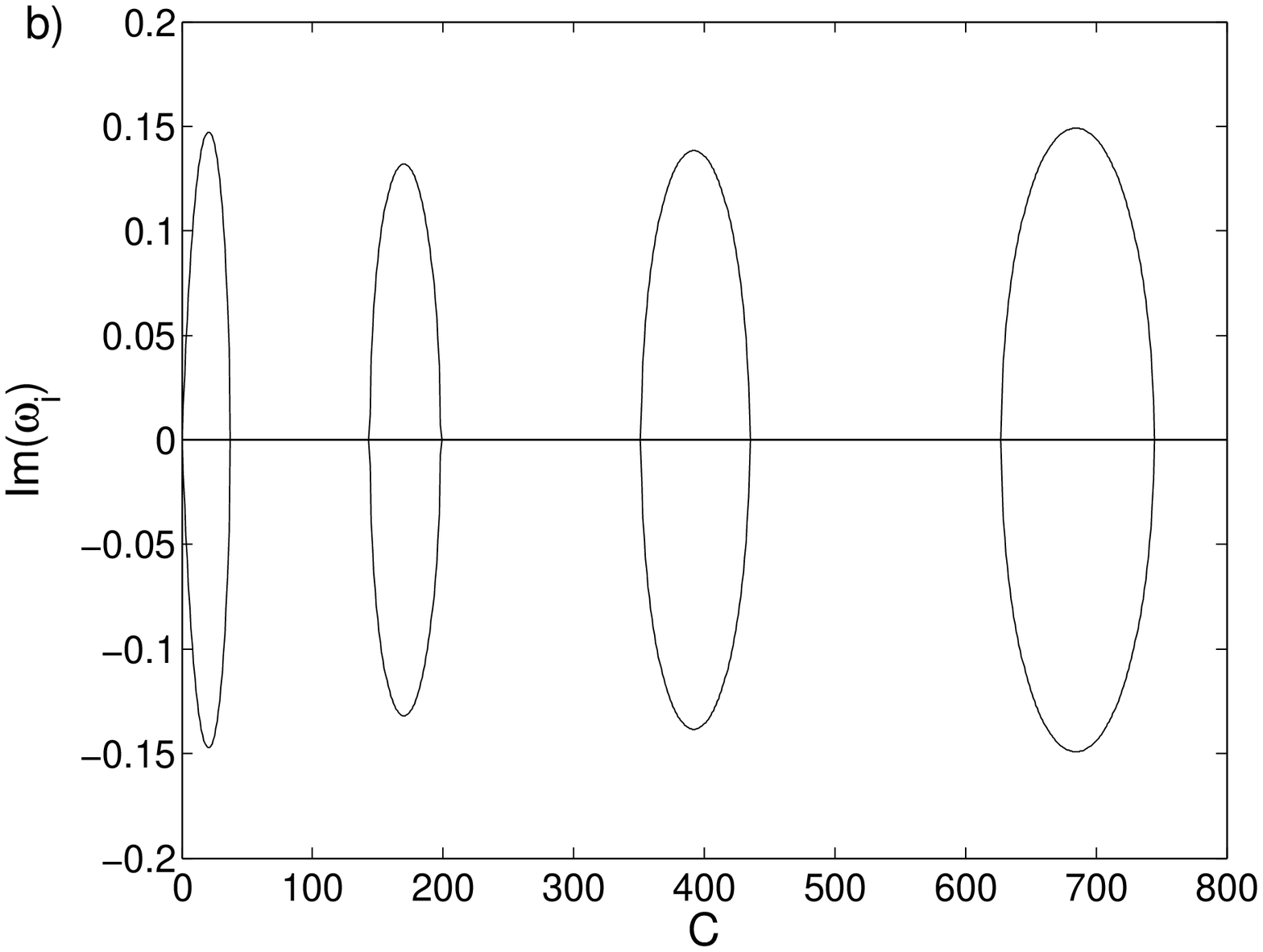}
\caption[]{Energy levels in two dimensions for a condensate with an $m=2$ 
  vortex. The left panel
  shows the real parts and right panel the imaginary parts. All imaginary 
  parts except at most one are zero at any point in this phase space.}
\label{fig:levels_2d}
\end{figure}

As a starting point for the study of the dynamics it is useful to
repeat the linear stability analysis \cite{pu1999,Lundh06}. The
GP equation is expanded about a stationary solution
$\Psi(\vec{r},t)=\Psi_0(\vec{r})\exp(-i\mu t)$ (which in the present 
case will be the doubly quantized vortex solution), where $\mu$ is the 
chemical potential of the system. The ansatz for the expansion is 
taken to be 
\beq
\label{psi_plus_bog}
\Psi(\rr,t) = \left[\Psi_0(\rr)+\sum_{n}\left(u_n(\rr) e^{-i\omega_n t} 
+ v_n(\rr)^* e^{i \omega_n t}\right)\right]e^{-i\mu t},
\enq
where $u_n$ and $v_n$ are the quasiparticle amplitudes
and $\omega_n$ the
quasiparticle energies calculated from the Bogoliubov equations 
\cite{Pethick2001}. 
The small-amplitude excitations of the 
condensate are described by the eigenvectors and eigenvalues of the 
Bogoliubov equations, 
\beq
\label{eq:bogoliubov}
B
\left(\begin{array}{l} u_n(\rr) \\ v_n(\rr)\end{array}\right)
= \omega_n\left(\begin{array}{l} u_n(\rr) \\ v_n(\rr)\end{array}\right),
\enq
where the linear 
operator $B$ is defined by
\bea
\label{eq:bogoliubovmat}
B = \left( \begin{array}{cc}
H_0-\mu+2C|\Psi|^2 & C\Psi^2 \\
- C(\Psi^*)^2 &-\left(
H_0 
-\mu + 2C|\Psi|^2 \right)
\end{array}\right).
\ena
If $B$ has a complex eigenvalue, the system 
is dynamically unstable and the corresponding mode will grow 
exponentially. It is known that there exist intervals of the 
coupling constant $C$ where the Bogoliubov equations 
possess a pair of complex eigenvalues. 
This behavior was thoroughly studied by the present authors in a
previous paper \cite{Lundh06} (cf.~\cite{pu1999}). 
Figure \ref{fig:levels_2d} shows the
eigenvalue behavior as a function of $C$ for the 2D case.
An instability occurs when the energies of two Bogoliubov modes 
collide. In the present case the mode confined to the interior of 
the vortex, referred to as the core mode, mixes with surface modes 
of quadrupole symmetry. The core mode is seen in Fig.\ 
\ref{fig:levels_2d}(a) as the line with positive 
slope that repeatedly merges with other lines representing the energies 
of quadrupole surface modes; each such collision 
creates an instability, 
so that successively higher instability regions correspond to 
increasing radial quantum number of the quadrupole mode.

The Bogoliubov equations describe only the linear, i.~e., 
small-amplitude, evolution of the condensate. In order to 
capture the full, nonlinear time development, in general 
one has to perform a numerical time integration 
of the time-dependent GP equation (\ref{eqn:GPE}). However, 
the purpose of the present paper is to study to what extent a 
simplified approach, based on the solutions to the Bogoliubov 
equations, will suffice, and therefore we will in subsequent 
sections compare the full numerical results to the simplified model. 
To study the splitting dynamics of a doubly quantized vortex 
one needs to choose a perturbed doubly quantized vortex as initial
condition. The doubly quantized vortex state is a stationary, 
rotationally symmetric solution of the GP equation (\ref{eq:dimlessgpe}) 
of the form 
\beq
\Psi_2(r,\theta)=f_2(r)e^{i 2 \theta}, 
\enq
where the real amplitude $f_2(r)$ obeys the equation 
\begin{equation}
  \left[ -\frac{1}{2}\left(\frac{1}{r}\frac{\partial}{\partial r}
  r\frac{\partial}{\partial r} +\frac{2^2}{r^2}\right)+V(r)+C |f_2(r)|^2\right] f_2(r)=\mu
  f_2(r).\label{eq:2m_vortex}
\end{equation}
As the 
initial condition for dynamical simulations one needs to add a 
perturbation to the 
doubly quantized vortex state. For definiteness, we have chosen to 
use the ground-state
harmonic oscillator wave function as a perturbation, but as long as the 
perturbation is small, its exact form does not matter for the long-time 
evolution, after it is exponentially inflated.

\section{Numerical method}
\label{sec:numerical_method}
The GP equation (\ref{eq:dimlessgpe}) 
is solved using a Hermite mesh in both spatial 
directions, and the time evolution is done using a Strang splitting 
that makes use of 
the tensor product structure of the linear problem \cite{McPeake2002}. 
For a sufficiently large grid, in 
our case $100\times 100$ points, we get conservation of angular momentum to
one part in $10^{6}$. This symplectic method is nearly optimal for the
problem at hand, which was crucial in order to be able to scan the parameter
range and to analyze the subtle nonlinear dynamics in detail.

The Bogoliubov equation is solved separately. 
Due to the cylindrical symmetry, it is reduced to a 1D eigenvalue problem, 
whose solution was described in Ref.\ \cite{Lundh06}.

One of the most important quantities to be discussed in the following 
is the distance between two vortices in the numerical time evolution. 
To measure this distance, we first identify the spatial points 
$\rr$ which fulfill the criteria $|\rr|< 4$ and 
$|\Psi(\rr)|<0.15 {\rm Max}(\Psi)$; these are the points of low density 
in the interior of the system. 
Using these points we do a least-square fit to the form 
\begin{equation}
  \widetilde{\Psi}(z)=A(z+z_0)(z-z_0) e^{-|z|^2/2}, 
\label{eqn:fitting}
\end{equation}
where $z$ is short for $z=x+i y$. The fit is done with respect to the 
two constants $A$ and $z_0$. 
This fitting function describes two vortices 
placed symmetrically about the origin and is found to be an accurate 
approximation for the wavefunction at all times, in accordance with 
the expectation that the instability of a doubly quantized vortex results 
in the vortex splitting into two. 
The fit for the parameter $z_0$ gives the positions of the two vortices 
as $z_0$ and $-z_0$, and the
vortex distance is $d=2|z_0|$. A good fit is very difficult to achieve
for small separations, since the least-square method minimization problem
is then very shallow and small numerical errors in the wavefunction 
$\Psi$ give significant contributions. A more reliable method to find
the qualitative time evolution is to notice that in the weakly
interacting limit, the squared length $|z_0|^2$ is approximately
proportional to the population of the
lowest harmonic-oscillator eigenstate
(see \cite{Lundh06}, Eq.~(23)). Therefore we project 
the wave function onto the eigenstates of the harmonic-oscillator
potential,
\begin{equation}
  a_{n,m}(t)=\int \Psi(x,y,t) \phi_{n,m}(x,y) dx dy,
\label{eq:project}
\end{equation}
where $\phi_{n,m}$ is the eigenstate of the harmonic-oscillator 
potential with energy $\omega_{n,m}=2n+|m|+1$,
\beq
\phi_{n,m}(r,\theta) = 
\sqrt{\frac{n!}{\pi(n+m)!}}L_{nm}(r^2)\left(re^{i\theta}\right)^m
e^{-r^2/2},
\label{eqn:ho_states}
\enq
and 
\beq
L_{n\alpha}(x)=\sum_{j=0}^{n}(-1)^j \binom{n+\alpha}{n-j} 
\frac{1}{j!}x^j
\enq
is a generalized Laguerre polynomial. 
The population of an excited 
state is defined as 
\beq
P_{0,m}(t) = |a_{0,m}(t)|^2.
\enq
The integral in Eq.\ (\ref{eq:project}) is calculated using
the Gauss-Hermite quadrature rule associated with the Hermite mesh,
which is exact in the limit of low energies. As we shall se, we 
find the amplitudes $P_{n,m}$ useful for understanding the
dynamics of the problem.

\section{Time development of vortex splitting}\label{sec:nummerical_calculations}
As known from previous studies \cite{pu1999,Lundh06},
the dynamics of a perturbed
doubly quantized vortex falls into one of two categories depending on the
value of the coupling strength $C$. 
In some intervals the doubly quantized vortex is stable and
in others it is unstable, as investigated in detail in Ref.\
\cite{Lundh06}. The real and imaginary parts of the Bogoliubov eigenvalues 
are presented in Fig.\ \ref{fig:levels_2d}. 
The regions where the vortex is linearly 
stable are not interesting from a dynamical perspective when small
perturbations are considered. The condensate will just perform small 
periodic oscillations following the initial perturbation. 
Thus the domains of interest are the unstable regions. It turns out 
that these can roughly be divided into two: the first unstable region, 
and all the subsequent ones. 

\begin{figure}[ht]
\includegraphics[width=0.4\textwidth]{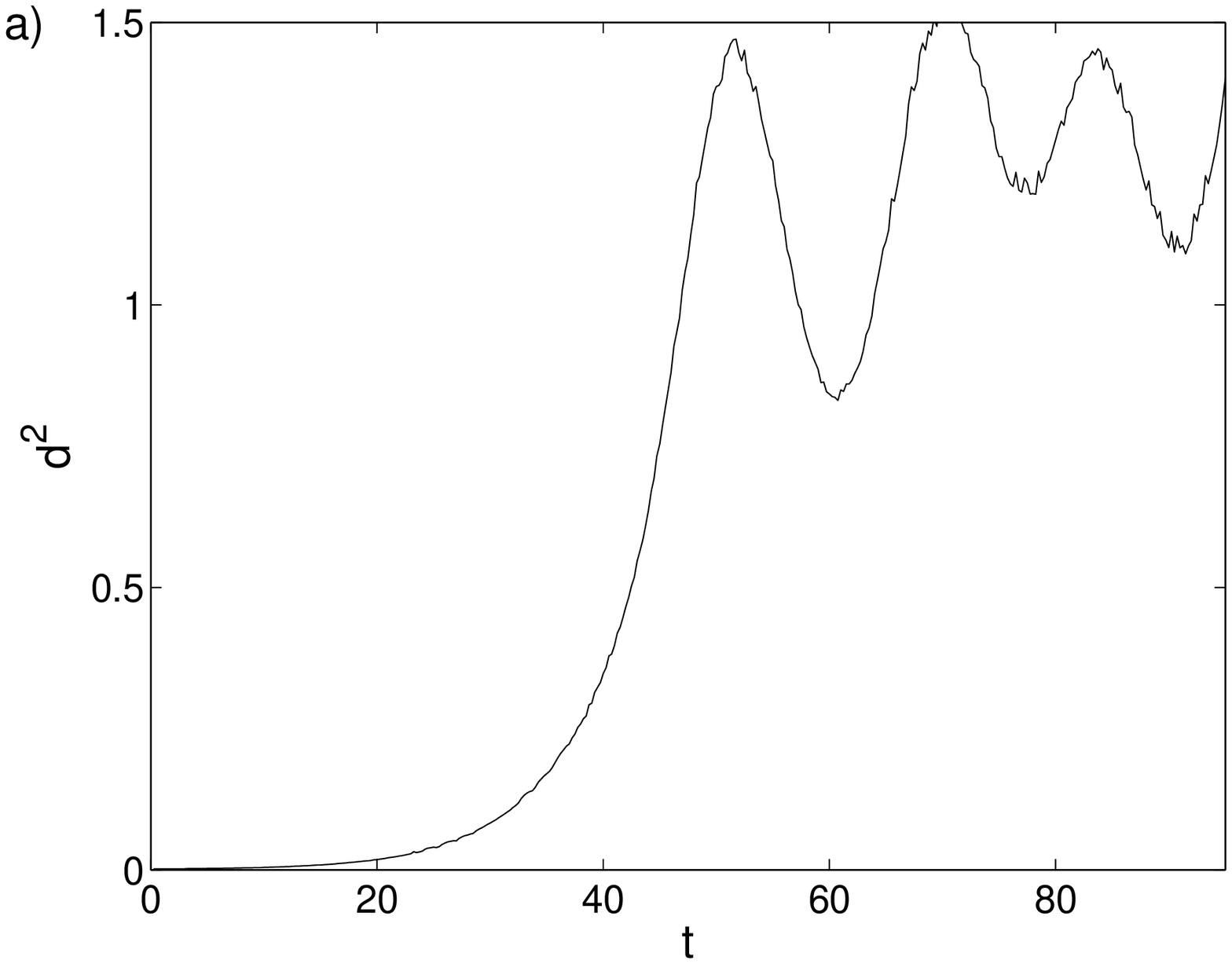}
\includegraphics[width=0.4\textwidth]{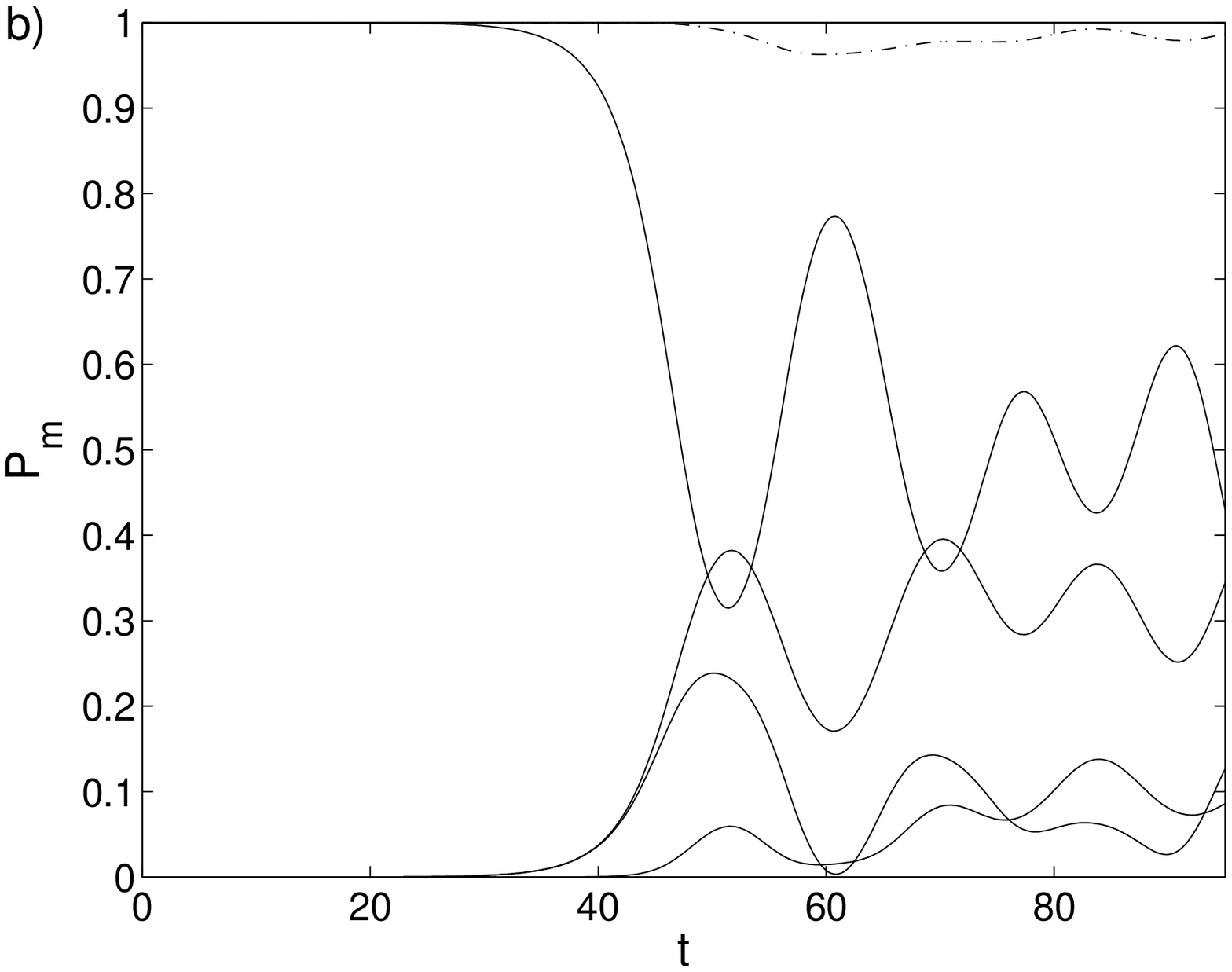}
\caption[]{(a) Time development of the vortex distance $d$, and  
  (b) time development of the populations $P_{m0}(t)$ of the 
  harmonic-oscillator eigenstates, for 
  a two-dimensional condensate with coupling parameter 
  $C=20$. In (b), the curves represent from the bottom up, 
  $m=6$, $m=4$, $m=0$, $m=2$, and the sum of all four. 
  The initial state was perturbed by means of a seeding of the 
  harmonic oscillator ground state with an amplitude $P_{00}=0.001$.}
\label{fig:C20_prob_vortex_dist}
\end{figure}

We first consider the first unstable region, $C\in [0,37]$. 
An example of the dynamics 
is given in Fig.\ \ref{fig:C20_prob_vortex_dist}. The 
depletion of the condensate, i.~e.\ the $m=2$ state, is very strong. 
It is seen that the sum of the populations in the $m=0$, 2, and 4 
states is less than 1 after some time, which means that there is 
a non-negligible population in states with $m>4$. (Although the 
negligible population of states with odd $m$ is here a consequence of the 
chosen initial conditions, we have checked that for more general 
initial conditions it is enforced by the dynamics, since 
only modes with even $m$ become dynamically unstable.)
The population in states with $m>4$
is seeded by the large population in the $m=4$ state, as will 
be clear below. Another feature which is worth 
noticing is that the vortex distance is highly correlated with the $m=0$
population, as anticipated in Sec.\ \ref{sec:nummerical_calculations}. 
We take advantage of this near proportionality to find the time 
dependence of the
vortex distance when the fitting method to find the vortex position 
described in Sec.\ \ref{sec:nummerical_calculations} 
fails.

The time evolution proceeds in two stages. From the start the
population of the $m=0$ state (which is the perturbation inserted by hand) 
and the $m=4$ state grow exponentially
while the condensate, the $m=2$ state, is accordingly depleted. 
After the population of the $m=0$ and $m=4$ states has become non-negligible,
the populations of the two amplified states becomes asymmetric, due 
to population of higher-angular momentum states. 
The vortex
distance and the population will start oscillating around finite
values. Later we will see that the asymmetry and the population of
higher-angular momentum eigenstates are crucial for the vortex distance to
not oscillate back to zero. It is important to note that the
asymmetry between the $m=0$ and $m=4$ populations is not caused by the initial
population chosen here, but is enforced by the dynamics. 

\begin{figure}[ht]
\includegraphics[width=0.4\columnwidth]{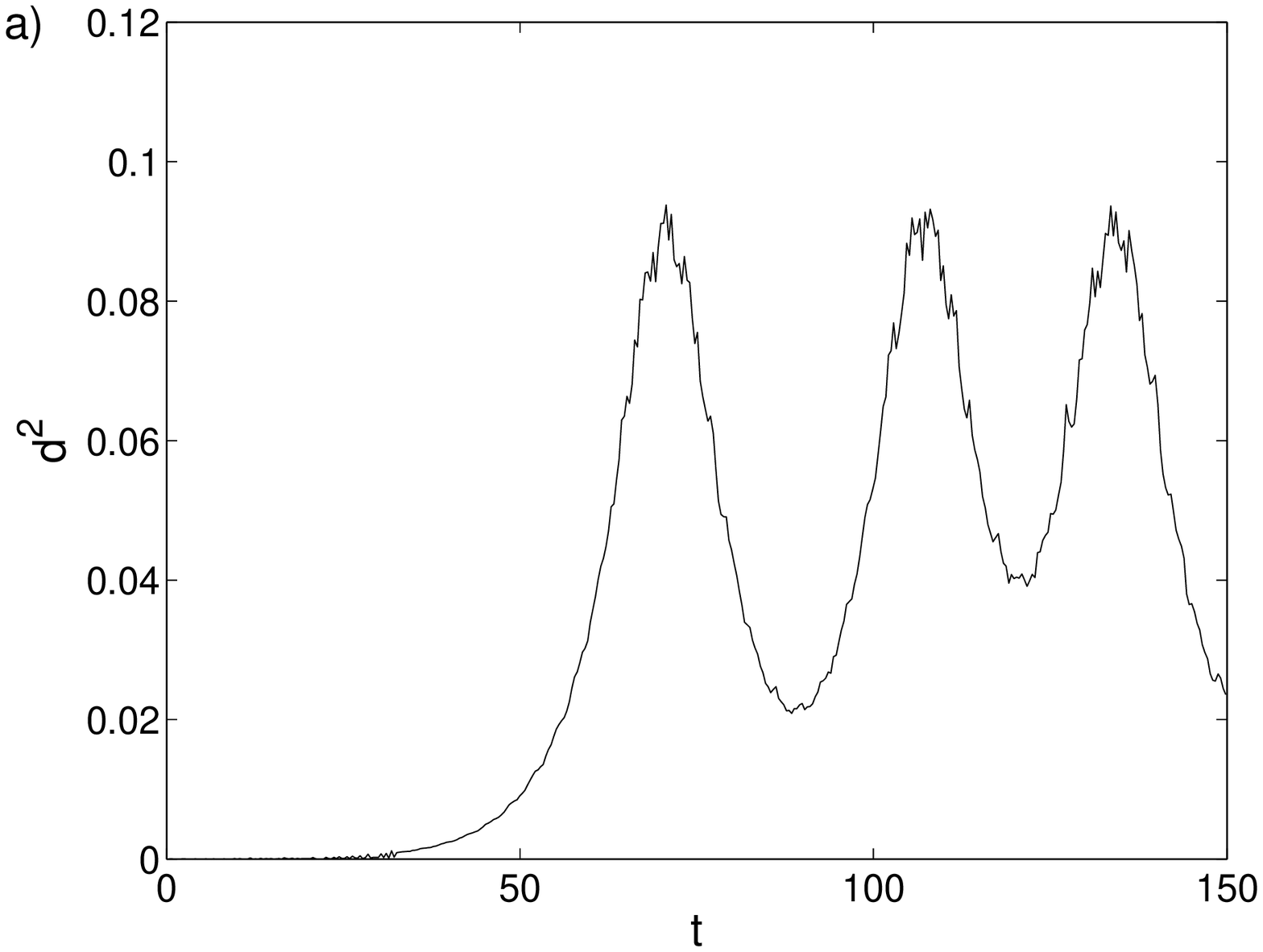}
\includegraphics[width=0.4\columnwidth]{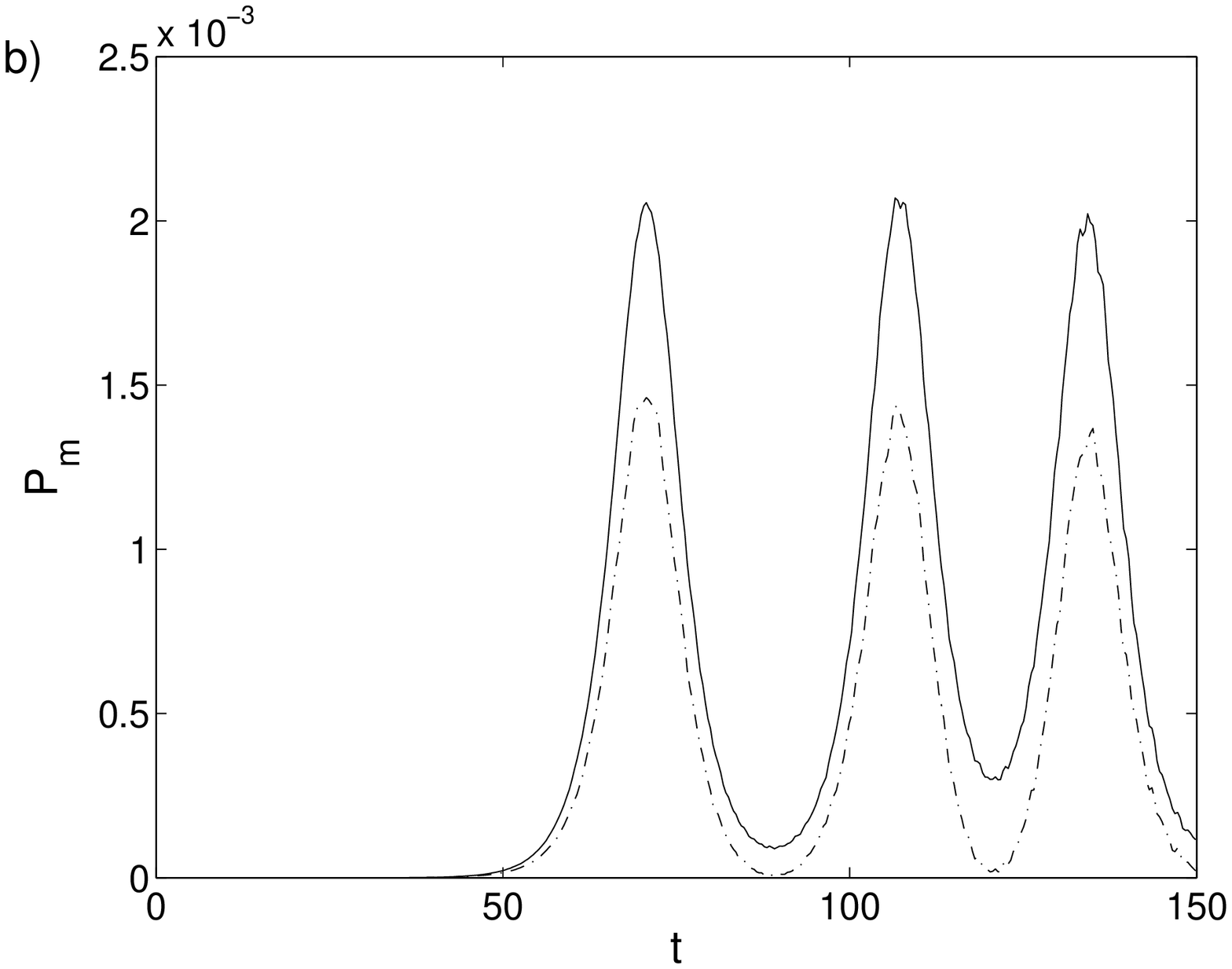}
\caption[]{(a) Square of the vortex distance $d$, and 
  (b) total population in the $m=0$ (upper curve) and $m=4$ 
  (lower curve) states, 
  for a condensate when the coupling 
  parameter $C=380$. 
  The initial seeding of of the harmonic-oscillator ground state is
  $P_{00}=1\times10^{-5}$.}
\label{fig:prob_C_380}
\end{figure}

The dynamics in the higher unstable regions is different from that in 
the first. Figure
\ref{fig:prob_C_380} plots the 
population in different $m$ states for 
$C=380$, which is located in the third instability region 
(see Fig.\ \ref{fig:levels_2d}).  
Like in the first instability region, the time evolution 
of the unstable
modes starts with an exponential growth. It achieves a maximum and start
to oscillate. In contrast to the small-$C$ case the oscillation is
dominated by one frequency. 
Furthermore, it is seen in Fig.~\ref{fig:prob_C_380} that 
the excited-state populations are very small at all times, and so is the
depletion of the condensate. This is a general feature of the time 
evolution of higher instability regions, and it will enable us to make 
a simple model that captures the main features of the vortex 
dynamics and at the same time is analytically solvable (see
Sec.\ \ref{sec:vortex_dynamic_model}). 
Finally, it is seen that the inter-vortex distance shows the same time 
dependence as the mode population $P_{0,0}$. The maximum distance 
between the two vortices is about $d=1$ (in units of the oscillator 
length $\aos$ as always), which is similar to that in the first unstable 
region, but contrary to that case, the diameter of the condensate is 
now much larger, meaning that the two vortices will stay well inside the
condensate. The vortices rotate around each other and the 
distance between them oscillates in a non-sinusoidal way. 

\begin{figure}
\includegraphics[width=\columnwidth]{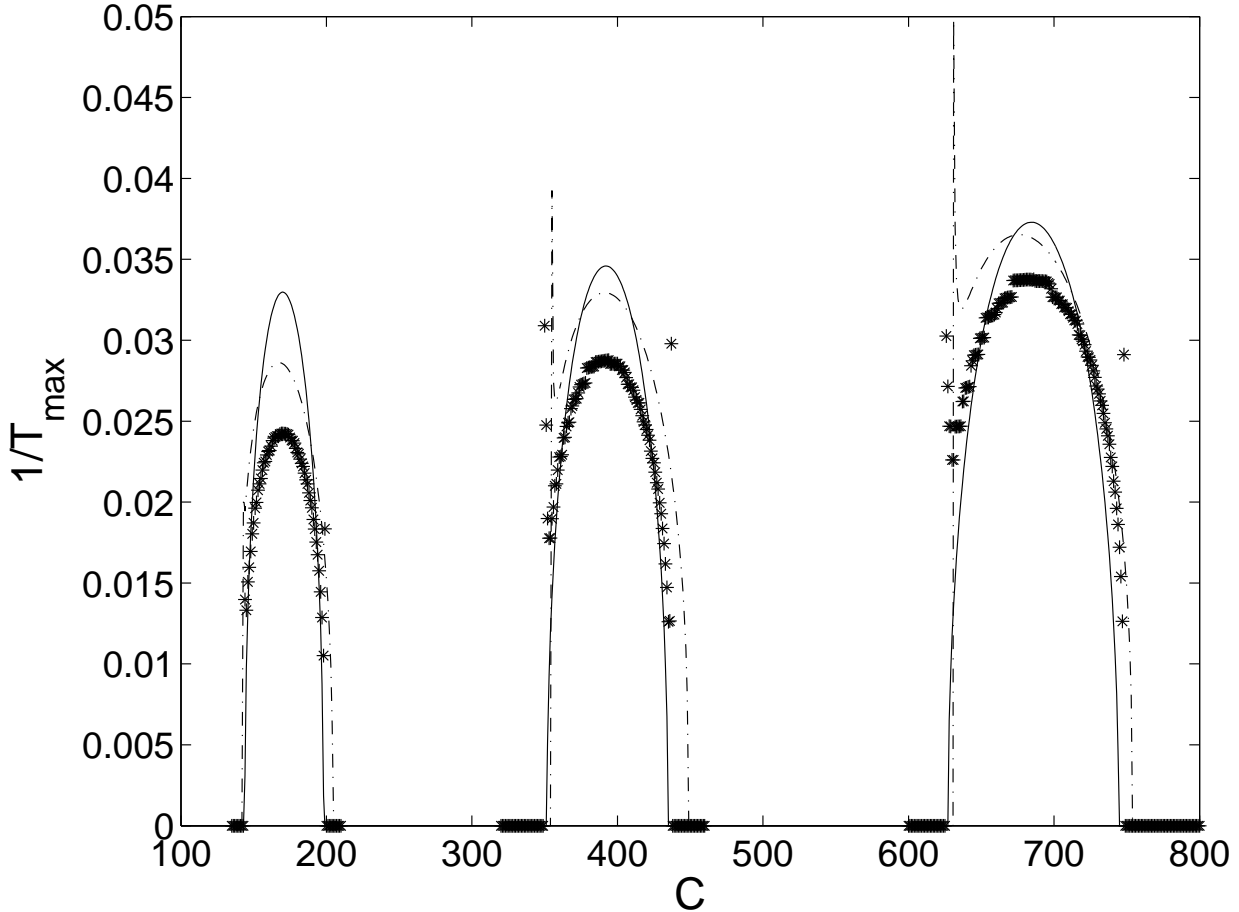}
\caption[]{Time for splitting of a doubly quantized vortex. Full lines 
represent the imaginary part of the complex Bogoliubov eigenvalue,
asterisks represent the time $T_{\rm max}$ taken for the vortex distance to 
achieve its first maximum according to the full GPE solution, 
and the dashed line is the same time scale in the three-state model, 
Eq.\ (\ref{maxtime}).}
\label{fig:omega_Tmax}
\end{figure}

From the discussion above, we may identify the most important 
characteristics of 
the dynamics as follows: (i) the exponential growth factor, (ii) the time 
until the first
maximum is achieved, (iii) 
the maximum of the amplitude of the excited state, and (iv) the 
maximum inter-vortex distance.
All of these features are
functions of the nonlinear parameter $C$ only. 
It is seen that items (i) and (ii) are closely related. The growth factor 
is given by the largest complex part of the Bogoliubov eigenvalues, 
while the time until first maximum must be inferred from numerical 
calculations; a comparison of these two is shown in 
Fig.\ \ref{fig:omega_Tmax}. 
The dashed line in Fig.\ \ref{fig:omega_Tmax} is the result of the 
three-state model that will be described in Sec.\ \ref{sec:vortex_dynamic_model} below.
We see that in all instability regions the imaginary part of the 
mode frequency 
agrees well with the inverse of $T_{\rm max}$. 


\begin{figure}[ht]
\includegraphics[width=0.4\columnwidth]{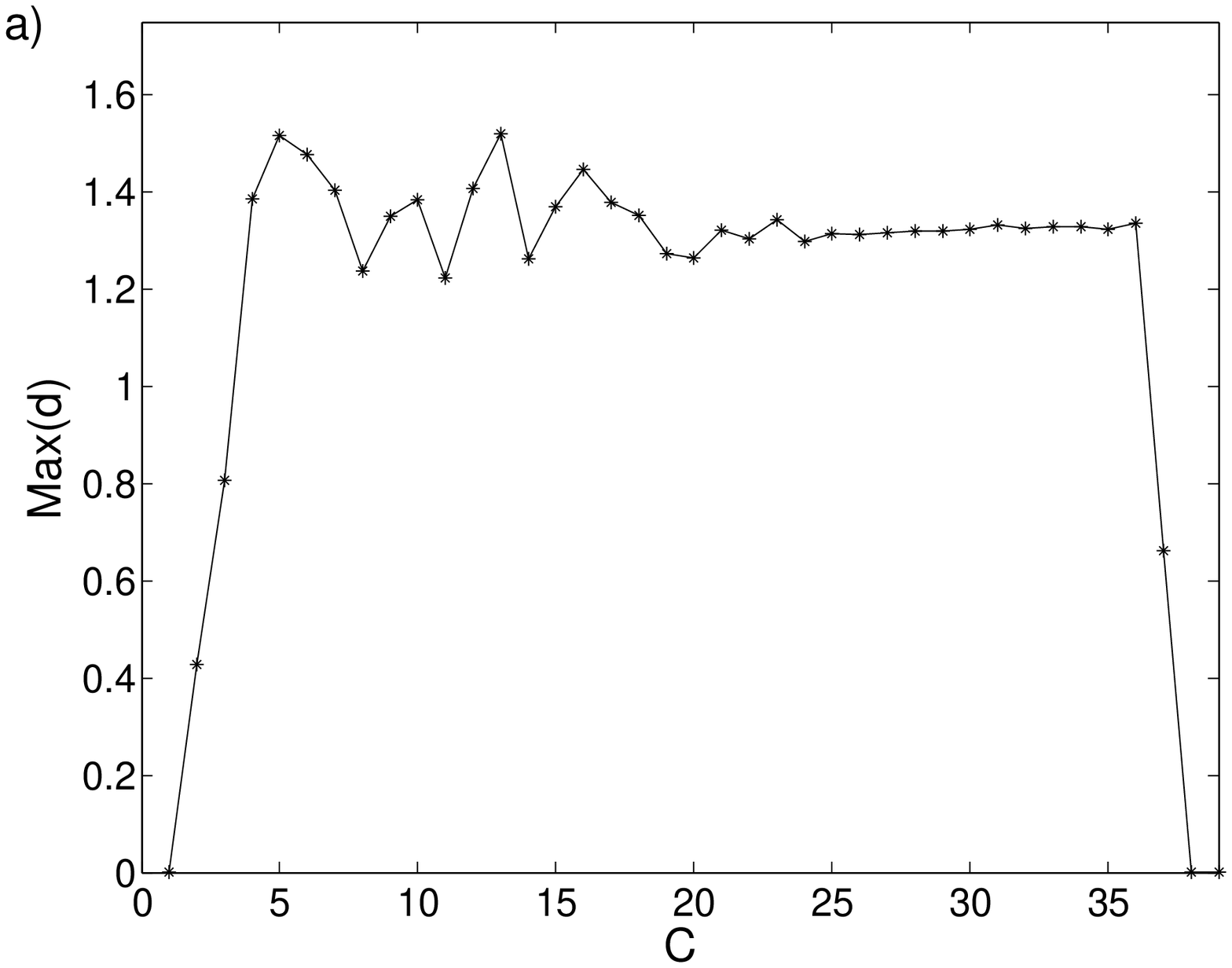}
\hspace{0.06\columnwidth}
\includegraphics[width=0.4\columnwidth]{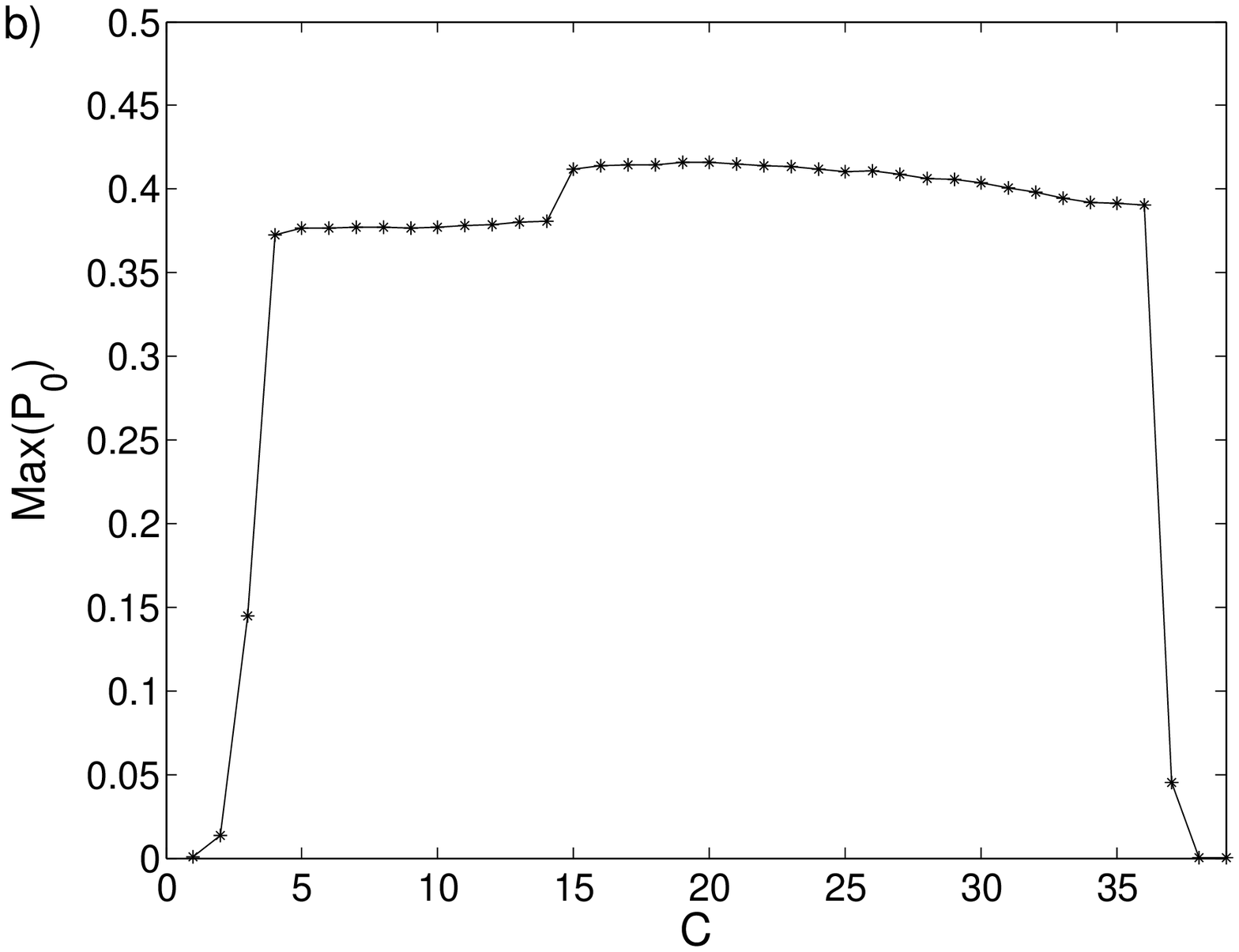}
\caption[]{(a) The maximum radius of the motion of the vortices, and
  (b) the maximum occupation of the $m=0$ state, as a function of 
  coupling parameter C
  with initial seeding $P_0(0)=0.01$. The chosen range of $C$ values 
  lies in the
  first instability region.}
\label{fig:Max1_39}
\end{figure}

\begin{figure}[ht]
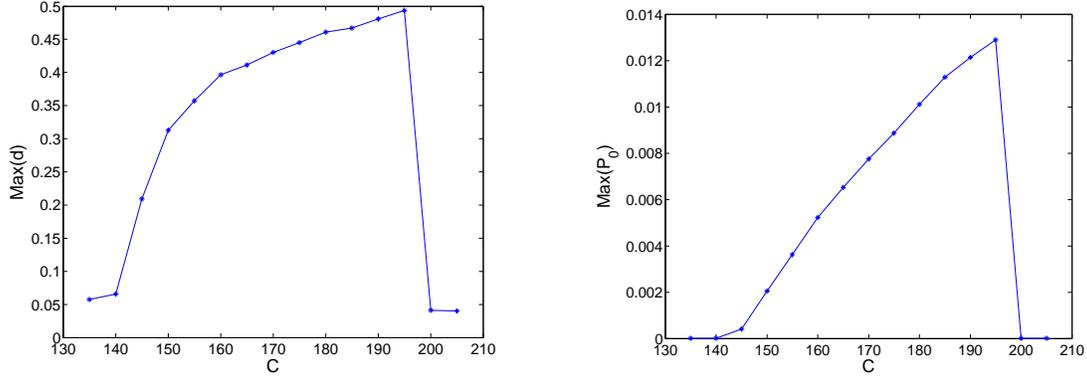

\includegraphics[width=0.4\columnwidth]{DATA_scan_tverr_135_205_liten_0.001MaxVdist.epsc}
\hspace{0.06\columnwidth}
\includegraphics[width=0.4\columnwidth]{DATA_scan_tverr_135_205_liten_0.001MaxM.epsc}
\caption[]{(a) The maximum radius of the motion of the vortices, and (b)
  the maximum probability of the $m=0$ state, $P_0$, as a function of 
  coupling parameter $C$
  with initial seeding $P_0(0)=0.001$. The chosen range of $C$ values lies 
  in the second instability region.}
\label{fig:Max135_205}
\end{figure}

On the other hand, items (iii) and (iv), the maximum amplitude of 
the vortex distance and the
maximum population of the unstable mode, 
show a quite surprising behavior. For the first
instability region we see in Fig.\ \ref{fig:Max1_39} that both the
maximum of the vortex distance and the maximum of the population are
approximately independent of $C$ in the unstable region. 
This is despite that the
time to achieve this maximum varies strongly with $C$.

For the higher unstable regions the behavior is very different. We 
present the result for the region $C\in[135,205]$ in Fig.\ 
\ref{fig:Max135_205}. The behavior of the maximum amplitude is 
particularly interesting since it has an almost linear increase from
the start of the unstable region and reaches a maximum at the 
strong-coupling end of the unstable region, where it jumps 
discontinuously to zero. This behavior will be explained in Sec.\ 
\ref{sec:vortex_dynamic_model}, where it is also shown that  
just outside this discontinuity a finite-amplitude
perturbation can bring the system into the unstable region where the
amplitude grows approximately to the maximum value achieved at the
discontinuity. Finally we note in Fig.~\ref{fig:Max135_205} that 
the maximal vortex distance $d$ displays a
similar behavior if one takes into account that $P_0\sim d^2$.

\section{Three-mode dynamics of vortex splitting}
\label{sec:vortex_dynamic_model}
This section is devoted to extracting the main features of the dynamics
of the splitting dynamics of the doubly quantized vortex that was 
studied numerically in the previous section. We put up a nonlinear 
model that can be solved analytically and which 
captures the main dynamics of the full system. The parameters of the 
model can be extracted from the the GP equation, for the most part 
analytically, and are all functions
of $C$. This model will be particularly accurate for the higher
unstable regions.

\subsection{First instability region}

It is instructive to first consider the dynamics in the first
unstable region \cite{kavoulakis2004}. To find approximative solutions to the 
the GP equation, it is useful to start from the 
Lagrangian 
from which the full GP equation can be derived if 
no further approximations are invoked \cite{Pethick2001},
\begin{equation}
    L = i\int d\rr \frac{1}{2}(\hat{\Psi} \frac{\partial \Psi}{\partial t}
  -\Psi \frac{\partial \hat{\Psi}}{\partial t})
  -(\hat{\Psi} H_0 \Psi+\frac{C}{2}|\Psi|^4).
\label{eqn:full_lagrangian}
\end{equation}
In the limit of small $C$, we know that the the dynamics mainly involves
three states, namely the lowest-energy harmonic-oscillator eigenstates 
$\phi_{n,m}=\phi_{0,0}$, $\phi_{0,2}$, and $\phi_{0,4}$ 
(see Eq.~\ref{eqn:ho_states}). 
The $m=2$ state represents the condensate, and $m=0$ and $m=4$
the core and surface states respectively, which will be populated 
due to the instability.
To investigate the dynamics of the vortex splitting in the space 
spanned by the three states, we expand the wave function as
\begin{equation}
  \Psi(\rr,t) = a_0(t) \phi_{0,0}(\rr) + a_2(t) \phi_{0,2} +a_4(t) \phi_{0,4},
\end{equation}
so that $a_m$ is the amplitude of the state with $m$ quanta of
angular momentum in the $z$-direction.
If we insert this into the Lagrangian 
we obtain 
\begin{eqnarray}
  L&=& i\sum_m\left(a_m^*\dot{a}_m -\dot{a}^*_m a_m \right)\nonumber\\
  &-&  \sum_m \omega_m a_m^* a_m 
  - \frac{1}{2} \sum_m C_{m,m}(|a_m|)^4 \nonumber\\
  &-& \sum_{m<m'}C_{m,m'}|a_m|^2|a_{m'}|^2 \nonumber\\
  &-&  K (a_0
  {a_{2}^{2}}^*
   a_4+a^*_0 a_{2}^{2} a^*_4)
\end{eqnarray}
where $C_{m,m'}= C \int |\phi_{0,m}|^2|\phi_{0,m'}|^2 d^2r$, with $m,m'=0,2,4$, and
$K= C \int |\phi_{0,0}| |\phi_{0,2}|^2 |\phi_{0,4}| d^2r$. The angular momentum
conservation is automatically taken care of by the symmetries of the 
eigenfunctions. 
Using this expression we can write down the nonlinear equations of 
motion 
\begin{eqnarray}
  i \dot{a}_0 &=& (E_0+C_{0,0} |a_0|^2+ C_{0,2} |a_2|^2+ C_{0,4} |a_4|^2)
  a_0 +K a_2^2 a^*_4\\
    i \dot{a}_2 &=& (E_2+ C_{2,0} |a_0|^2+C_{2,2} |a_2|^2+ C_{2,4} |a_4|^2)
  a_2 +2 K a_0 a^*_2 a_4\\
  i \dot{a}_4 &=& (E_0+C_{4,0} |a_0|^2+ C_{4,2} |a_2|^2+ C_{4,4} |a_4|^2)
  a_4 +K a_0^* a_2^2. \\\label{eq:trunc_model}
\end{eqnarray}
Making a variable change $a_m \to \widetilde{a}_m = a_m e^{i\theta_m(t)}$, 
with a suitable choice of phases $\theta_m$, 
the system of equations can be rewritten as 
\begin{eqnarray}
  i \dot{\widetilde{a}}_0 &=& K \widetilde{a}_2^2 \widetilde{a}^*_4e^{i\phi(t)}\nonumber\\
    i \dot{\widetilde{a}}_2 &=& 2 K \widetilde{a}_0 \widetilde{a}^*_2 \widetilde{a}_4e^{-i\phi(t)}\nonumber\\
  i \dot{\widetilde{a}}_4 &=& K \widetilde{a}^*_0
    \widetilde{a}_2^2 e^{i\phi(t)}, 
\label{eq:trunc_model_phas}
\end{eqnarray}
where the phase is 
\begin{eqnarray}
  \phi(t)&=&\int^{t}
(E_0+C_{0,0} |a_0|^2+ C_{0,2} |a_2|^2+ C_{0,4} |a_4|^2)\\
  &+&(E_4+C_{4,0} |a_0|^2+ C_{4,2} |a_2|^2+ C_{4,4} |a_4|^2)\\
  &-&2(E_2+ C_{2,0} |a_0|^2+C_{2,2} |a_2|^2+ C_{2,4} |a_4|^2) dt.\label{eqn:trunk_model}
\end{eqnarray}
It can be seen from Eq.\ (\ref{eq:trunc_model_phas}), that if initially 
$\widetilde{a_0}(0)=\widetilde{a_4}(0)$, then it holds that 
$\widetilde{a_0}(0)=\widetilde{a_4}(0)$ at all times. 

This equation is expected to give the correct dynamics for small $C$, 
where the dynamics is dominated by the lowest-energy single-particle 
eigenfunctions. For larger $C$ values, the structure of the equation 
is still expected to be the same. As we will see below, the $m=2$ 
wavefunction should then
be replaced by the condensate wavefunction and the $m=0$ and $m=4$ 
states with the
two Bogoliubov states associated with core excitations and surface 
excitations, respectively.

\begin{figure}[ht]
\includegraphics[width=\columnwidth]{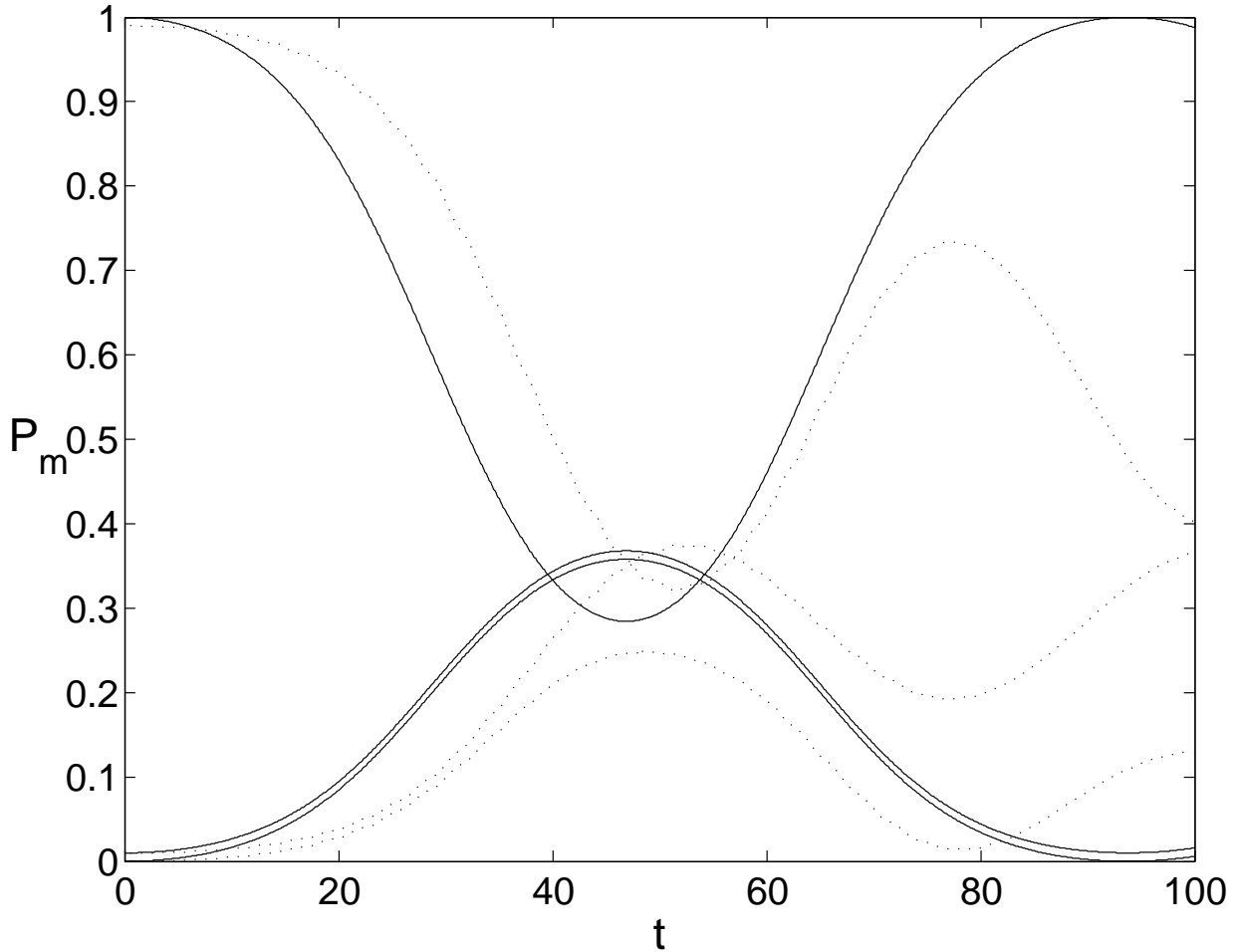}
\caption[]{The time dependence of the amplitude of the lowest $m$
  states in the truncated three-state model in equation
  \ref{eqn:trunk_model}, compared to the full GPE simulation.
  The interaction strength is $C=5$ and the initial seeding $P_0=0.1$.
  From top to bottom, the full lines represent the amplitudes $P_2$, 
  $P_0$, and $P_4$ in the truncated model and the dotted lines represent 
  the corresponding amplitudes for the full GPE simulation.}
\label{fig:comp_model}
\end{figure}
In Fig.\ \ref{fig:comp_model} we see that as long as the population 
is concentrated to the three states used in the truncation, the
evolution of the truncated equation is identical to the full
solution shown in Fig.\ \ref{fig:C20_prob_vortex_dist}. 
The main cause for discrepancies is that the $m=6$ state 
starts to become populated, which causes a relative depletion of the 
$m=4$ state. This results in an asymmetry between
the populations of the $m=4$ and $m=0$ states, and also implies that 
the amplitude $a_0$ (and with it the vortex distance) will not 
return back to zero, as it does in the truncated 
model. 
The terms in the Lagrangian 
(\ref{eqn:full_lagrangian}) causing this conversion are of the form 
$a_2a_4a_0^*a_6^*$ and $a_4^2 a_2^*a_6^*$, and thus they are 
proportional to three powers of the excited-state populations. 
In situations where the population of all higher states is small,
the population of $m=6$ and higher states is expected to be much slower, 
and this is also seen in Fig.\ \ref{fig:prob_C_380}; thus a three-state 
model should be more accurate for higher instability regions 
than for the first.

The truncated system has six degrees of freedom, corresponding to the 
real and imaginary parts of the three amplitudes $a_m$, but it has to 
conserve energy, norm and angular momentum, which leaves three degrees 
of freedom. In addition, the relative phase of the coefficients 
$\widetilde{a_0}$ and $\widetilde{a_4}$ will according to 
Eq.~(\ref{eq:trunc_model_phas}) stay constant; thus the 
system has only two degrees of freedom, which makes it integrable, 
so that the solution is periodic. 

The temporal dynamics depends on the initial state; since the initial 
increase of the excited-state population is exponential, it is expected 
that the time taken to achieve the first maximum is proportional 
to the logarithm of the initial population of the excited states. It is 
checked numerically that this logarithmic behavior is in fact very accurate 
even for the full nonlinear evolution.


\subsection{Higher instability regions}
For the dynamics in the higher unstable regions we have to modify 
the three-state model so that it 
takes into account the energy of the condensate and the
coupling dependence of the quasiparticle energies. The structure of
the model should also be such that it conserves angular momentum,
quasiparticles and energy. The ansatz is therefore written 
\beq
\Psi(\rr,t) = e^{i\mu t} \left[ 
  a_2(t)\Psi_2(\rr) + a_0(t)u_0(\rr) + a_0^*(t)v_0^*(\rr)
  + a_4(t)u_4(\rr) + a_4^*(t)v_4^*(\rr)\right],
\label{eq:ansats}
\enq
where $\Psi_2(\rr)$ is the condensate wavefunction with a doubly 
quantized vortex, and 
$u_0$ and $v_0$ are the Bogoliubov amplitudes 
for the core mode with $m=0$. Finally, $u_4$ and $v_4$ are the 
Bogoliubov amplitudes for a selected quadrupole mode, which is 
expected to become unstable when it 
mixes with the core mode. In Ref.\ \cite{Lundh06} it was found that 
an instability occurs when the energy of the $(n,m)=(0,0)$ Bogoliubov 
mode, the core mode, becomes nearly degenerate with a quadrupole mode 
with quantum numbers $(n,m)=(n,m)$; the recurring instability regions 
arise from the crossings with quadrupole modes with 
successively higher $n$ values; this can be seen in Fig.\ 
\ref{fig:levels_2d}. 
All the functions in Eq.~(\ref{eq:ansats}) are assumed 
to be calculated from Eqs.\ (\ref{eq:dimlessgpe},\ref{eq:bogoliubov}) 
at some fixed coupling strength $C$ outside of any instability region; 
their energies are then to be extrapolated into the instability 
region. 

The calculations are carried out in Appendix \ref{app:utleding}. 
As already noted, the energies of the two Bogoliubov modes 
are nearly degenerate, and are assumed to coincide at a coupling $C_0$. 
Furthermore, 
since the core mode is concentrated to the interior of the vortex, its 
energy varies much more rapidly with coupling strength $C$ than that 
of the quadrupole mode, so that only the $C$ dependence 
of the former needs to be taken into account. Again, this is 
seen in Fig.\ \ref{fig:levels_2d}. Moreover, the same 
confinement also leads to a self-interaction of the core mode; 
corresponding terms for the other modes are small in comparison. 
Putting all this together results in the coupled equations 
\begin{eqnarray}
  i \dot{a}_0 &=& (2dE - 2I_0 |a_0|^2)
  a_0 +K a_2^2 a^*_4\nonumber\\
    i \dot{a}_2 &=&2 K a_0 a^*_2 a_4\nonumber\\
  i \dot{a}_4 &=& K a^*_0 a_2^2.
  \label{eqn:coupled}
\end{eqnarray}
Here, $dE=(\omega_0-\omega_4)/2$ is half the $C$-dependent energy 
difference between the Bogoliubov eigenenergies;  
at the resonant coupling strength $C_0$ we have 
$2dE(C_0)=\omega_0(C_0)-\omega_4=0$, so we may write 
\begin{eqnarray}
  dE
  = \frac12\frac{\partial \omega_0}{\partial C}(C-C_0).
\end{eqnarray} 
Inserting the 2D Thomas-Fermi approximation \cite{Pethick2001}
$\mu(C)=(C/\pi)^{1/2}$, and using the expression for the core mode energy 
\cite{Lundh06}, $\omega=0.42 \mu$, we obtain 
\begin{equation}
  dE = 0.42 \frac{1}{2\sqrt{\pi C_0}}(C-C_0).
\label{eq:w04_TF}
\end{equation}
The term $I_0$ in Eq.~(\ref{eqn:coupled}) represents the 
nonlinear self-interaction of the core mode, 
\begin{equation}
  I_0=\frac{1}{2}C \int d\rr |v_{0}|^4(\rr) \approx 
  \frac{C_{0}^{3/2}}{4\sqrt{3} \pi^{3/2}},
\label{eq:dw_TF}
\end{equation}
where the last equality was carried out in App.\ \ref{app:utleding}.
With Thomas-Fermi estimates for the core mode frequency \cite{Lundh06},
\beq
\omega_0(C) = 0.42\sqrt{\frac{C}{\pi}},
\enq
and the quadrupole mode frequency for the $n$'th radially 
excited state \cite{stringari},
\beq
\omega_4 = \sqrt{2n^2+6n+2},
\enq
the resonant coupling $C_0$ was obtained in Ref.\ \cite{Lundh06} 
as 
\beq
C_0 = \frac{\pi}{0.42^2}(2n^2+6n+2),
\label{eq:C0}
\enq
where each value of $n$ corresponds to an instability region. 
Finally, the constant $K$ 
represents the integral that couples the three modes; it is found that any 
attempt to approximate this term analytically is extremely sensitive to 
small variations in the variational parameters, so $K$ has to be 
determined 
numerically. This can be done by noting (as will be shown in a moment) 
that the constant is fact equal to the maximum 
of the imaginary parts of the mode frequencies over the instability 
interval; numerically it is seen to be close to $K\approx 0.15$ for 
all instability regions. We note that $I_0$ is at least an order 
of magnitude larger than $K$ when $C$ is of order 100 or more; this 
inequality will be taken advantage of in the calculations. Also note 
that whereas $K$ and $I_0$ are positive as long as the 
interactions are repulsive, the sign of $dE$ depends on $C-C_0$.

To see that $K$ is related to the maximum imaginary part of the 
mode frequency, linearize Eq.\ (\ref{eqn:coupled}) by removing the 
term proportional to $I_0$ and put $|a_2|=1$; the resulting 
oscillating solution for the amplitudes $a_0$ and $a_4$ has a 
frequency 
\beq
\omega_{\rm lin} = dE \pm
\sqrt{dE^2-K^2}
\label{eq:linear_freq}
\enq
in accordance with Bogoliubov theory. From this we conclude that the 
mode is unstable when $|K/dE| < 1$, and that $K$ is indeed 
the maximum imaginary part of the frequency.

In Appendix \ref{app:full_nl_model} it is shown how the system of 
equations (\ref{eqn:coupled}) leads to the differential equation 
for the core mode population $p=|a_0|^2$, 
\begin{equation}
  \dot{p}^2=
  -\left[2(dE-K)p-(I_0-4K) p^2-E_0\right]
   \left[2(dE+K)p-(I_0+4K) p^2-E_0\right]
  \equiv f(p),
  \label{eq:P_eq}
\end{equation}
where the constant $E_0$ is the total energy. 
A formal solution is
\begin{equation}
  \int_{p(0)}^{p(t)}\frac{{\rm d}p}{\sqrt{f(p)}}
  = t.
\end{equation}
This is an elliptic integral since 
$f(p)$  is a polynomial
of degree 4. The solution for $p(t)$ is therefore given as an inverse of this
elliptic integral. To understand the dynamics of the system we look at the
zeros of $f(p)$.
The solution will oscillate between the two positive roots of $f(p)$, 
since they correspond to $\dot{p}=0$. In the limit 
$E_0 \ll (K,dE) \ll I_0$ [which holds according 
to the discussion below Eq.\ (\ref{eq:C0})], the roots can be written
\bea
p_0 &=& \frac{E_0I_0}{2(K+dE)^2},\nonumber\\
p_{\mathrm max} &=& 2\frac{dE+K}{I_0}-p_0.
\label{eq:roots}
\ena
In a typical experimental situation, $p_0$ is the initial value.

In App. \ref{app:solve_model} it is found that the 
asymptotic expansion for the time to the first maximum, under the 
inequalities stated above, is given by 
\begin{eqnarray}
  T&\sim&\frac{1}{4\sqrt{K^2-dE^2}}\ln\left(\frac{16}{k'^2}\right), \nonumber\\
  k'^2&\approx&\frac{1}{4(1-(\frac{dE}{K})^2)^2}\frac{I_0^2}{
  (K+dE)^2}p_0^2.
\end{eqnarray}
The time scale is set by the imaginary part of the eigenvalue
of the linearized problem, $\sqrt{K^2-dE^2}$, as long as we 
have $|dE/K|<1$, as discussed in connection with 
Eq.\ (\ref{eq:linear_freq}). The
contribution of the initial population is only logarithmic. The 
nonlinearity described by the constant $I_0$ also contributes a 
logarithmic term. 
We conclude that the splitting time is mainly predicted by the
linear Bogoliubov theory, and the nonlinear dynamics contributes only weakly. 

\begin{figure}[ht]
\includegraphics[width=0.3\columnwidth]{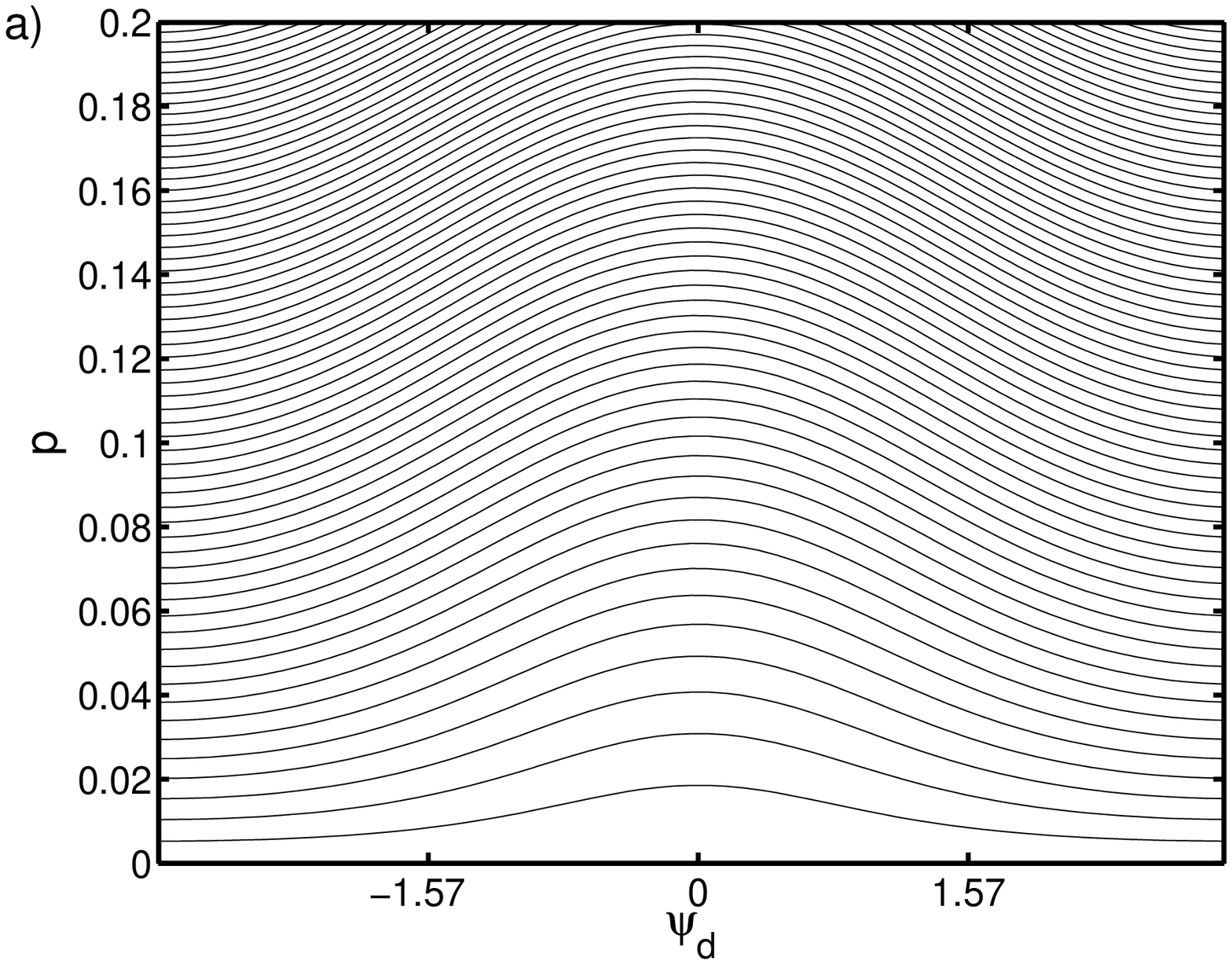}
\hspace{0.03\columnwidth}
\includegraphics[width=0.3\columnwidth]{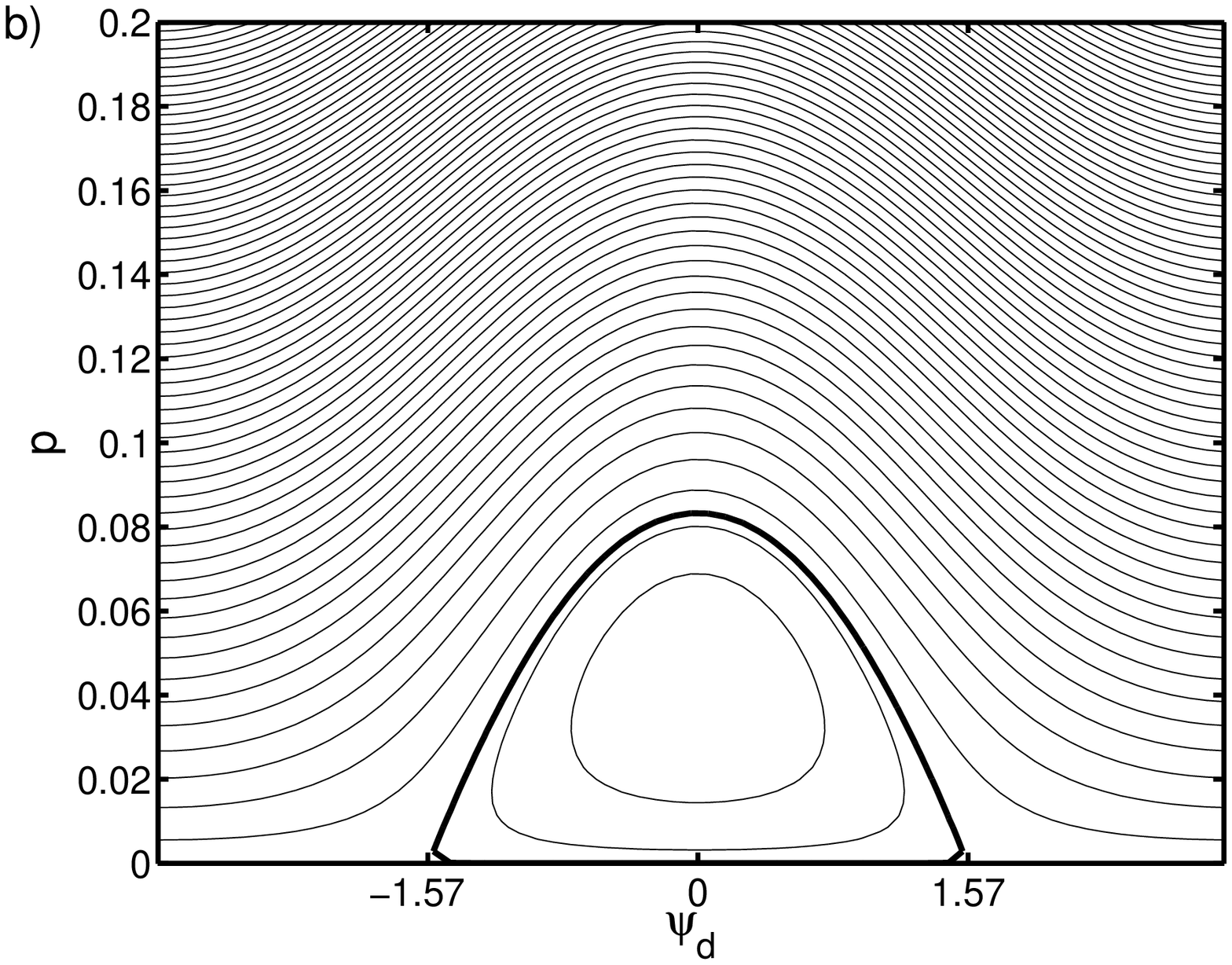}
\hspace{0.03\columnwidth}
\includegraphics[width=0.3\columnwidth]{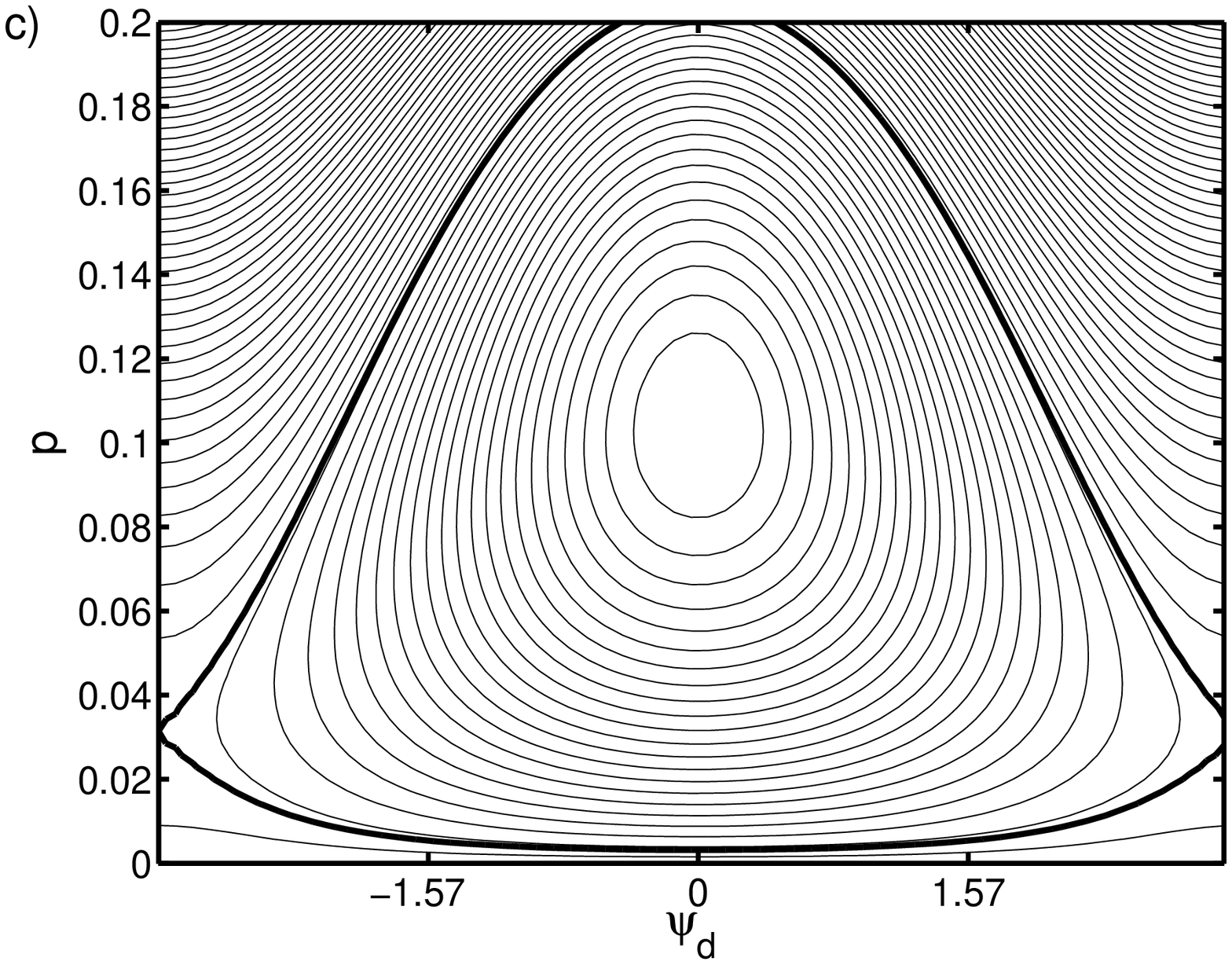}
\caption[]{Contour plots of the Hamiltonian (\ref{eqn:model_hamiltonian}) 
  for
  $dE=-1.5$ (a), $dE=0$ (b), and $dE=1.5$ 
  (c). The other parameters are chosen as $I_0=10$ and $K=1$. 
  The thick 
  line is the separatrix, dividing phase space into running and oscillating 
  solutions.}
\label{fig:contour_hamiltonian}
\end{figure}

To further understand the dynamics is is useful to examine the
Hamiltonian associated with these equations of motion. It is given by 
Eq.\ (\ref{eqn:model_hamiltonian}) as  
\begin{equation}
  H(\psi_d,p)=2K(\frac12-p)p\cos\psi_d+(2dE-I_0 p) p,
\end{equation}
where $\psi_d$ is twice the phase difference between the two modes $a_2$ and 
$a_0$, $\psi_d=2\psi_2-2\psi_0$.
Again, we can according to Eqs.\ 
(\ref{eq:w04_TF}-\ref{eq:dw_TF}) take the physically motivated 
limit $K/I_0\ll 1$.
In Fig.\ \ref{fig:contour_hamiltonian} we see a contour plot of
the Hamiltonian. 
Fig.\ \ref{fig:contour_hamiltonian}(a) shows the case 
$dE<-K$, where 
$p=0$ is a global maximum and the Hamiltonian is strictly
convex, i.e., $p=0$ is unconditionally stable. In the linearly
unstable regime, $K>|dE|$ [Fig.\ \ref{fig:contour_hamiltonian}(b)], 
the point $p=0$ is a saddle point and the Hamiltonian
has a global maximum for $p=(dE+K)/I_0,\psi_d=0$. 
The thick 
line in Fig.\ \ref{fig:contour_hamiltonian} is the 
separatrix, separating the solutions where $\psi_d$ oscillates from the 
running ones. 
Finally, 
for $dE>K$, as shown in (c), 
$p=0$ is a local minimum 
and the Hamiltonian has a saddle point at
$p=(dE-K)/I_0$ and $\psi_d=\pi$. In this case the solution 
is stable
when the initial conditions are sufficiently near $p=0$; else it may
start to oscillate around the maximum.

The zeros of the
polynomial $f(p)$ defined in Eq.\ (\ref{eq:P_eq}) 
correspond to points where the tangent of a 
contour line is horizontal. We observe that the contours are of 
two different types depending on whether they are closed lines, that do 
not wind about the origin, or whether they wind around phase space and 
connect at $\psi_d=\pm\pi$. If the initial condition is purely
imaginary, $\psi_d=\pi$,  then the solution will always lie on a curve 
that winds around phase space. 
In the case $\psi_d=0$ the solution can lie on any level curve depending 
on the initial
condition. 
Consider the case where $dE>K$, i.e., $C$ is above the unstable 
region. Then a small initial value of $p$ will yield a solution that lies 
in the stable region, i.e. the solution 
will circle around the local maximum. However, if the initial value of 
$p$ is increased, the system enters a trajectory that winds around the 
minimum and $p$ starts to oscillate. This is the reason
for the finite-amplitude instabilities above the upper limit of the
unstable region that were observed numerically in 
Sec.\ \ref{sec:nummerical_calculations}.

\begin{figure}[ht]
\includegraphics[width=0.4\textwidth]{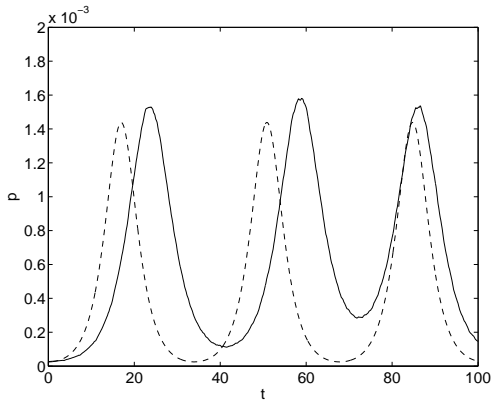}
  \caption{Time development of the population $p$ of the core mode. 
    Full line represents the full numerical time integration, while 
    the dashed line is obtained from the three-state approximation. 
    The coupling strength is chosen as $C=380$.} 
\label{fig:comp1}
\end{figure}

\begin{figure}[ht]
\includegraphics[width=\columnwidth]{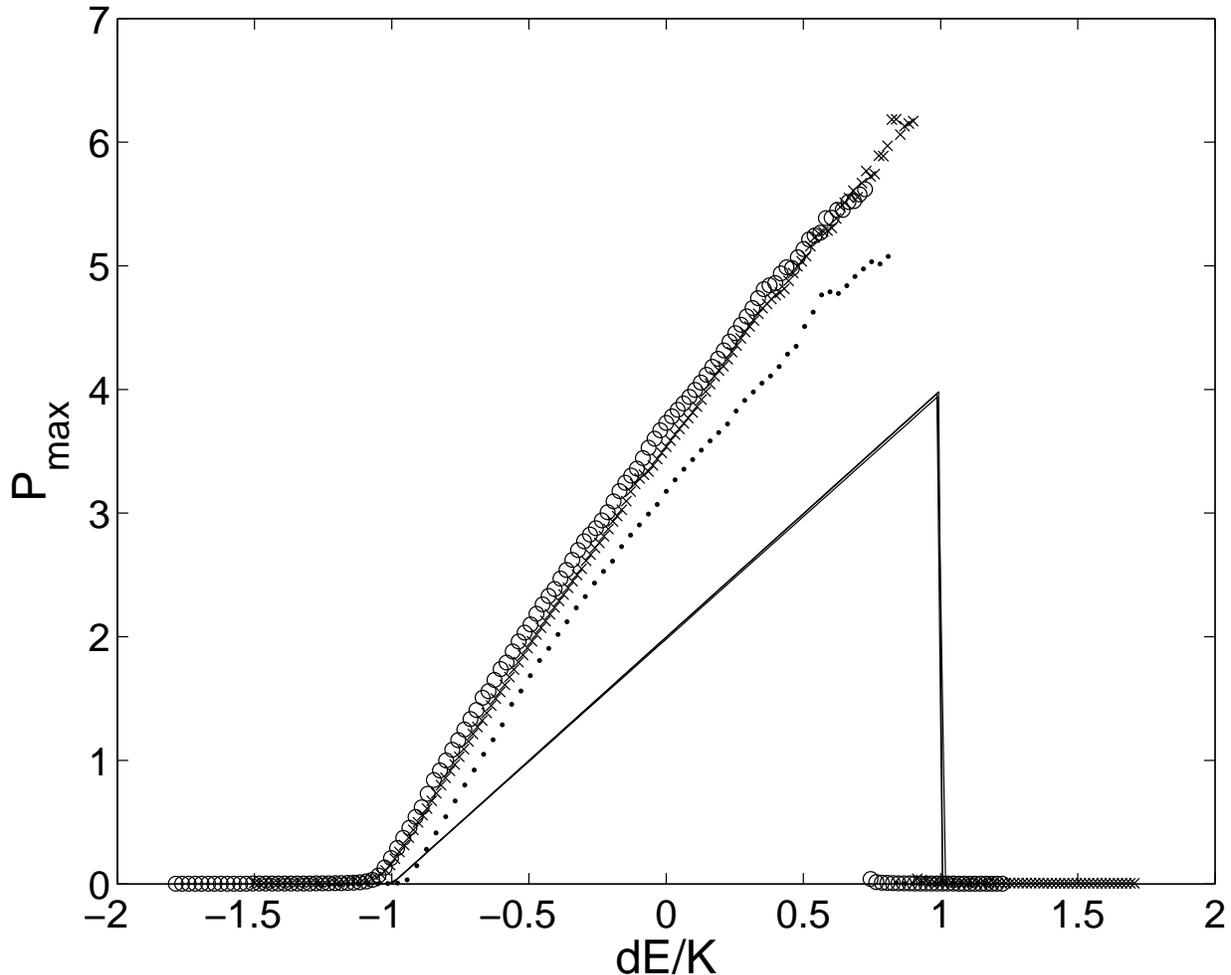}  
  \caption{Maximum amplitude $p_0$ of the core mode as a function of 
  energy difference divided by mode coupling, $dE/K$. The full line 
  is the analytical solution of the three-state model, and the symbols 
  represent data from the GPE solution in successive instability 
  regions; circles for the second instability region, crosses for the 
  third, and dots for the fourth.}
\label{fig:comp2}
\end{figure}

In Figs.\ \ref{fig:comp1}-\ref{fig:comp2} and Fig.\ \ref{fig:omega_Tmax} 
we compare the three-state 
model calculation with the full time integration.
The overlap $p$ is in the numerical calculation defined as the 
overlap integral between the condensate wave function and the core 
mode, analogously to Eq.\ (\ref{eq:project}), but with 
the numerically calculated core mode $v_0$ 
substituted for 
the single-particle eigenfunction $\phi_{0,0}$. 
Fig.\ \ref{fig:comp2} collapses numerical data from the solution of the 
GP equation in several instability regions onto the same graph, by 
for each instability region mapping the coupling parameter $C$ onto 
the parameters $dE$ and $K$. 
As seen, the agreement 
with the three-state model for the time $T$ in 
Fig.\ \ref{fig:omega_Tmax} is excellent, but the 
magnitude of the amplitude $p$, seen in Fig.\ \ref{fig:comp2},
is more sensitive to the exact parameter 
values, which may explain the discrepancy by a factor of order unity. 
Again, $p$ does not oscillate back to zero in the full solution, but 
it does in the truncated model. Clearly, the population of higher states 
makes the dynamics nonperiodic, so that the two vortices stay apart after they 
have separated.





\section{Conclusions}\label{sec:conclusion}
This study is concerned with the compressible vortex dynamics in a 
trapped Bose-Einstein condensate. The dynamics of the splitting of a 
doubly quantized vortex is studied in detail both in full numerical 
time integration, linear Bogoliubov analysis, and using a three-state 
model utilizing the Bogoliubov eigenstates. 
It is found that the simple three-state model captures many essential 
features of the dynamics. Moreover, it is seen that Bogoliubov analysis 
is capable of determining the time scale for vortex splitting, while 
nonlinear processes only contribute logarithmically.

\section{Acknowledgment}
Part of this work was supported by the Swedish research council, 
Vetenskapsr{\aa}det.

\appendix
\section{Derivation of three-state model}\label{app:utleding}
We start from the ansatz
\beq
\Psi(\rr,t) = e^{-i\mu t} \left[ 
  a_2(t)\Psi_2(\rr) + a_0(t)u_0(\rr) + a_0^*(t)v_0^*(\rr)
  + a_4(t)u_4(\rr) + a_4^*(t)v_4^*(\rr)\right],
\label{eq:utled_ansats}
\enq
where $u_m, v_m$ are the exact Bogoliubov amplitudes associated 
with the stationary condensate wave function $\Psi_2$ 
computed for a nonlinearity parameter $C$. 
$C$ is assumed to lie outside of all instability regions, and 
the energies will be extrapolated into them.
The dimensionless units 
were discussed in connection with Eq.~(\ref{dimensionless}).
We assume that the core state can be approximated well as a pure
hole state. This assumption amounts to putting $u_0=0$. The 
functions thus fulfill the equations 
\begin{eqnarray}
\int d\rr \Psi^*_2(\rr)  
(H_0 + C |\Psi_2(\rr)|^2)\Psi_2(\rr)&=&\mu, \nonumber\\
\int d\rr v^*_0(\rr) (H_0 +2 C |\Psi_2(\rr)|^2-\mu) v_0(\rr) &=& 
-\omega_0(C),\nonumber\\
\int d\rr u^*_4(\rr) [(H_0 +2 C |\Psi_2(\rr)|^2-\mu) u_4(\rr)+
C \Psi_2^2(\rr) v_4(\rr)]- \nonumber\\
\int d\rr v^{*}_{4}(\rr) [(H_{0} +2 C |\Psi_2(\rr)|^2-\mu) v_4(\rr)+C (\Psi^{*}_{2})^{2}
u_4(\rr)]&=&\omega_4(C).
\label{eq:eigeneqs024}
\ena
With the chosen sign conventions, $\omega_0$ and $\omega_4$ are both 
positive. In the absence of instabilities, $a_4$ is expected to 
oscillate as $\exp(-i\omega_4 t)$, and $a_0$ oscillates as 
$\exp(-i\omega_0 t)$. 
Taking nonlinearities into account, there will of course be
corrections to the time dependence. 
Also note the dependencies on the azimuthal angle $\theta$: 
$\Psi_2\propto\exp(2i\theta)$, $u_4\propto\exp(4i\theta)$, but $v_0$ and 
$v_4$ do not depend on $\theta$. 

On inserting the ansatz (\ref{eq:utled_ansats}) into the Lagrangian (\ref{eqn:full_lagrangian}), 
it separates into five parts,
\beq
L = L_k + L_0 + L_r + L_2 + L_4,
\enq
where $L_k$ contains the time derivatives,
\bea
  L_k&=& \frac{i}{2} (\conjg{a}_0\dot{a}_0 + \conjg{a}_2 \dot{a}_2 
  +\conjg{a}_4 \dot{a}_4 -c.c.) \nonumber\\
  &&+ \frac{i}{2}(a_0 \dot{a_4}\int d\rr v_0(\rr) \conjg{v}_4(\rr) -c.c.).
\ena
The term $L_0$ contains the terms to which 
the eigenvalue equations (\ref{eq:eigeneqs024}) can be applied,
\begin{eqnarray}
  L_0 &=&  -|a_2|^2\int d\rr\left[ \mu |\Psi_2(\rr)|^2 - 
    \Psi^*_2(\rr) (H_0 + C |\Psi_2(\rr)|^2) \Psi_2(\rr)\right] \nonumber\\
  && -|a_0|^2 \int d\rr v_0(\rr) (H_0 + 2 C |\Psi_2(\rr)|^2-\mu) v^*_0(\rr)
  \nonumber\\
  && -|a_4|^2 \int d\rr \left[u^*_4(\rr) (H_0 + 2 C |\Psi_2(\rr)|^2-\mu) u_4(\rr) 
    \right.\nonumber\\
  && \left.+v_4(\rr) (H_0 + 2 C |\Psi_2(\rr)|^2-\mu) v^*_4(\rr)\right],
\ena
whereas $L_r$ collects the ``rest terms'' obtained from $L_0$ because of  
the depletion of the condensate, 
\bea
  L_r &=& -C\,|a_2|^2(\frac{|a_2|^2}{2}-1) \int d\rr |\Psi_2(\rr)|^4 \nonumber\\ 
  &&  -2C |a_0|^2 (|a_2|^2-1)\int d\rr |\Psi_2(\rr)|^2 |v_0(\rr)|^2 \nonumber\\
  &&   -2 C|a_4|^2 (|a_2|^2-1) \int d\rr |\Psi_2(\rr)|^2 |u_4(\rr)|^2 \nonumber\\
  &&   -2 C|a_4|^2(|a_2|^2-1)\int d\rr |\Psi_2(\rr)|^2 |v_4(\rr)|^2.
  \label{eq:lr}
\ena
The term $L_2$ contains terms of second order in the excited-state 
amplitudes,
\begin{eqnarray}
L_2   &=&  -C [\conjg{a}_0 a^{*}_4 (a_2)^2 \int d\rr \Psi_2(\rr)^2 v^*_0(\rr) u^*_4(\rr)+c.c. ]\nonumber\\
  &&  -C a_0 a^{*}_4 |a_2|^2\int d\rr |\Psi_2(\rr)|^2 v_0(\rr) v_4(\rr)\nonumber\\ 
  &&-C [|a_4|^2|a_2|^2 \int d\rr\Psi_2(\rr)^2 u^*_4(\rr) v_4(\rr) + c.c.],
\ena
and analogously, $L_4$ contains the fourth-order terms,
\bea
L_4  &=&-\frac{C}{2}|a_0|^4 \int d\rr |v_0(\rr)|^4\nonumber\\
  &&-\frac{C}{2}|a_4|^2 \int d\rr (|u_4(\rr)|^4+ |v_4(\rr)|^4)\nonumber\\
  &&-C|a_0|^2 |a_4|^2 \int d\rr |v_0|^2(|v_4|^2+|u_4|^2) \nonumber\\
  &&-C|a_4|^4 \int d\rr |v_4(\rr)|^4|u_4(\rr)|^2. 
\label{eq:l4}
\end{eqnarray}
All the terms that are expected to oscillate as 
$\exp[i(\omega_0+\omega_4)t]$ are discarded. Next, consider all terms 
that are proportional to the fourth power of the excited-state amplitudes. 
Note that the function $v_0$ is concentrated to the 
vortex core, i.e., a very small spatial region, while the other functions, 
$\Psi_2$, $u_4$, and $v_4$, are much less localized. As a result, 
the term in the first line in Eq.\ (\ref{eq:l4}) is expected to be much 
larger than all the other terms in $L_4$, and those are therefore 
discarded. Furthermore, all the terms in $L_r$ are also of fourth order 
in the excited-state amplitudes and can be discarded. 
The resulting Lagrangian is 
\begin{eqnarray}
  L &=& \frac{i}{2}(\conjg{a}_0{\dot{a}}_0 + 
  \conjg{a}_4 \dot{a}_4 + \conjg{a}_2 \dot{a}_2 -c.c.)\nonumber\\
  &-& |a_0|^2 \omega_0(C)-|a_4|^2 \omega_4(C)\nonumber\\
  &-&   C\left(a_0 a_4 (a^*_2)^2 \int d\rr u_4 v_0 |\Psi_2|^2  + c.c.\right) \nonumber\\
  &-& \frac{C}{2} |a_0|^4 \int d\rr |v_0(\rr)|^4.
\end{eqnarray}

Defining the constants 
\bea
K =  C\int d\rr u_4 v_0 |\Psi_2|^2, \nonumber\\
I_0 = \frac{1}{2}C\int v_{0}^4, \nonumber\\
dE = \frac12[\omega_0(C)-\omega_4(C)], 
\ena
and making a phase change we can write the Lagrangian on the final form
\begin{eqnarray}
  L &=& \frac{i}{2}(\conjg{a}_0{\dot{a}}_0 + 
  \conjg{a}_4 \dot{a}_4 + \conjg{a}_2 \dot{a}_2 -c.c.)\nonumber\\
  &-& 2dE |a_0|^2 -  I_0 |a_0|^4
  -  \left( K a_0 a_4 (a^*_2)^2 + c.c.\right).
\end{eqnarray}

Following Ref.\ \cite{Lundh06}, one may produce Thomas-Fermi 
estimates for $I_0$ and $dE$, assuming that the core mode experiences 
an effective potential
\begin{equation}
  V(r)=2 C |\Psi|^2 = 2 \mu \frac{r^2}{\xi^2 b^2}
\end{equation}
where 
$\xi=1/\sqrt{2\mu}$ is
the healing length, and $b$ is a variational parameter;  
the choice $b=2\sqrt{6}$ minimizes the condensate energy.
The ground state of this potential is
\begin{equation}
  v_0(r)=\sqrt{\frac{\mu}{\pi\sqrt{3} }} \exp(-\frac{\mu}{\sqrt{3}} \frac{r^2}{2}),
\end{equation}
and the nonlinear parameter of our model becomes
\begin{equation}
  I_0 = \frac{C}{2} \frac{\mu}{\sqrt{3}} \frac{1}{2 \pi}
  = \frac{C^{3/2}}{4\sqrt{3}\pi^{3/2}},
\end{equation}
where in the last line we used the Thomas-Fermi result 
$\mu = (C/\pi)^{1/2}$.

\section{Solution of the coupled nonlinear system}\label{app:full_nl_model}
The Lagrangian for the system was in App.\ \ref{app:utleding} found to be 
\begin{equation}
  L({a_m},{a^*_m})=\sum_m \frac{i}{2}\left(a^{*}_m \dot{a}_m
  -{a_m}\dot{a}^*_m\right)-\left[(2dE - I_0 |a_0|^2) |a_0|^2
    +K \left((a_{2}^{*})^{2} a_0 a_4 +a_{2}^{2} a^{*}_{0} a^{*}_{4}
    \right)\right].
\end{equation}
First write the amplitudes in the form $a_i=r_i\exp(i\psi_i)$. The
Lagrangian can then be written in the form
\begin{eqnarray}
  L({r_m},{\psi_m})=\sum_{m} \dot{\psi}_m r^2_m 
  - \left\{2 K r_2^2r_0 r_4 \cos
  [2\psi_2-(\psi_0+\psi_4)]+
  (2dE-I_0 r_{0}^{2}) r_{0}^{2}\right\}.
\end{eqnarray}
Defining the auxiliary variables  
\begin{eqnarray}
    N&=&r_0^2+r_2^2+r_4^2, \quad
    \psi_n=2 \psi_2 -2 \psi_4,\nonumber\\
    L&=&2 r_2^2+4 r_4^2,
    \quad \psi_l=\frac{1}{4}( \psi_4 -  \psi_0),
    \nonumber\\
    D&=&r_0^2+\frac{1}{2}r_2^2+r_4^2,
    \quad \psi_d=-2 \psi_2 + \psi_0 + \psi_4,
\end{eqnarray}
the Lagrangian can be rewritten once more as 
\beq
  L(N,L,D,\psi_n,\psi_l,\psi_d) = 
  \dot{\psi}_n N+ \dot{\psi_l} L + \dot{\psi}_d D
  -\sqrt{F(D)}\cos\psi_d+G(D),
\enq
where
\bea
    F(D)&=&4 K^2  (2 N - 2 D)^2  (D - L/4) (-N + L/4 + D),\nonumber\\
    G(D)&=&[2dE -I_0(D - L/4)] (D - L/4).
\ena
We see that $N$ and $L$ are conserved; they are in fact the norm and 
angular momentum, respectively, so we suppress them as arguments. 
In addition the energy function
\begin{equation}
  E=\sqrt{F(D)}\cos\psi_d+G(D)
\end{equation}
is conserved.
The Lagrange equations for $D$ and $\psi_d$ read 
\begin{eqnarray}
  \dot{D}&=&-\sqrt{F(D)}\sin\psi_d,\nonumber\\
  \dot{\psi_d}&=&\frac{\partial}{\partial D} \sqrt{F(D)} \cos\psi_d
  + \frac{\partial}{\partial D} G(D). \label{eqn:genn_eqn}
\end{eqnarray}
Using energy conservation and the square of the first line we obtain 
the ODE
\begin{equation}
  \dot{D}^2=F(D)-(E-G(D))^2.
\label{eq:mellanresultat_d}
\end{equation}
This is an elliptic ODE since the right hand side is a polynomial of
degree $4$. The solution is done by factorizing the polynomial on 
the right-hand side into two second-order polynomials. 

Suppose $N=1$, $L=2$, which is the case in the present 
physical problem. In this case $F(D)$ is a square, 
\begin{equation}
  F(D)=f(D)^2\equiv 4 K^2 (1-D)^2(D-\frac12)^2,
\end{equation}
and the equation (\ref{eq:mellanresultat_d}) for $D$ simplifies to
\begin{eqnarray}
  \dot{D}^2&=&(f(D)+E-G(D))(f(D)-(E-G(D))).
\end{eqnarray}
Now rewrite the equation in terms of the original 
variable $p=r_0^2$ and obtain the final equation of motion, 
\begin{eqnarray}
   \dot{p}^2&=& P_1(p) P_2(p), {\rm where}\nonumber\\
  P_1(p)&=&(-4 K + I_0) p^2  + 2(K - dE) p + E,
  \nonumber\\
  P_2(p)&=&(-4 K - I_0) p^2  + 2(K + dE) p - E.
\label{eq:final_app_eq}
\end{eqnarray}
Note that the roots of the polynomial $P_1(p)P_2(p)$ may be either 
real or complex; two roots will become complex when 
$(dE-K)^2 < (I_0-4K) E$. Since $E$ is proportional to the 
initial population $p_0$ [see Eq.\ (\ref{eq:roots})], it can be 
assumed small and hence the complex roots appear only in a very small 
portion of phase space; this permits us to concentrate on the case with 
real roots only. Furthermore, as is shown in 
Sec.\ \ref{sec:vortex_dynamic_model}, 
we can on physical grounds assume $I_0 \gg K$; this will simplify 
some expressions in the following.

It is useful to write these equations as a Hamiltonian system with 
$\psi_d, p$ canonically conjugate variables. Starting from 
Eqs.\ (\ref{eqn:genn_eqn}) for $\psi_d,D$ and shifting variables as 
above to $\psi_d,p$ we obtain 
\begin{equation}
  H(\psi_d,p)=2K (\frac12-p) p \cos\psi_d + 
  (2dE-I_0 p) p.
\label{eqn:model_hamiltonian}
\end{equation}


\section{Elliptic integrals}\label{app:solve_model}
We now solve the differential equation for the core-mode amplitude 
$p$, Eq.\ (\ref{eq:final_app_eq}). Rewriting this as 
\beq
\dot{p} = I_0\sqrt{(p-\pmin)(p+p_1)(\pmax-p)(p_2+p)},
\enq
where $\pmin$ and $\pmax$ are the smallest and largest positive roots 
of the polynomial, respectively, and $-p_1$ and $-p_2$ are the other two, 
and taking the initial value for $p$ to be 
at the minimum point, then we may write 
\beq
  t= \frac{1}{I_0}
 \int_{\pmin}^{p(t)}\frac{dp}{[(p-\pmin)(p+p_{1})(\pmax-p)(p_{2}+p)]^{1/2}}.
\enq
Now define $p_l=(\pmin+p_1)/2$,$dl=(\pmin-p_1)/2$, $p_b=(\pmax+p_2)/2$,
$db=(\pmax-p_2)/2$, and $p'=p-dl$, to obtain the integral 
\beq
  t =
\frac{1}{I_0}
\int_{p_l}^{p'(t)}\frac{dp'}{[(p'^2-p_l^2)((p_b+db-dl)-p')((p_{b}-db-dl)+p')]^{1/2}}.
\enq
The asymmetry between the largest zeros
can be removed by invoking 
the substitution (\cite{Whittaker35}, p.\ 514)
\begin{equation}
  p'=\frac{a x+b}{c x+d}.
\end{equation}
The parameters are to be determined so that the transformation leaves 
the symmetric zeros of the integrand invariant but makes the other two
symmetric in terms of the new variable $x$; the new zeros of the denominator 
are denoted by $\pm p_s$. 
The integral for $t$ is now
\begin{eqnarray}
t&=& \frac{A}{I_0}\int_{p_l}^{x(t)}
\frac{dx}{\sqrt{(x^2-p_{l}^2)(p_{s}^2-x^2)}},
\ena
and assuming that all roots are real, as discussed in 
App.~\ref{app:full_nl_model}, the solution is 
\bea
\frac{I_0 p_s}{A} t 
&=& {\rm nd}^{-1}\left(\frac{x(t)}{p_l},1-\left(\frac{p_l}{p_s}\right)^2
\right),
\ena
where ${\rm nd}$ is a Jacobian elliptic function (\cite{Abramowitz72}, p.596), 
and
\beq
A^2 = \frac{(ad-cd)^2}{((a p_l)^2-c^2)(c-(a (p_b+dr)))(c+a (p_b-dr))},
\enq
with $dr=db-dl$. This gives the complete result
\begin{eqnarray}
 x(t)&=&p_{l}\, {\rm nd}( \frac{I_0 p_s}{A} t),\nonumber\\
 p(t)&=&\frac{a x(t)+b}{c x(t)+d} + dl.
\label{eq:full_solution}
\end{eqnarray}
The half period $T$ of the function $p(t)$ is given by the complete elliptic
integral 
\begin{equation}
T=\frac{A}{I_0 p_s}{\cal K}(k'^2),\quad k'^2=(\frac{p_l}{p_s})^2.
\end{equation}

Expanding the parameters in powers of $p_l$, which in our physical situation 
corresponds to assuming that 
the initial population is very small, yields
\begin{eqnarray}
A^2&\approx&
(1-(\frac{dr}{p_b})^2)-\frac{dr^2}{p_b^2}\frac{1}{(1-\frac{dr}{p_b})}\frac{p_l^2}{p_b^2}+{\mathcal O}(p_l^4).
\label{eq:c_param}
\end{eqnarray}
The transformation is given by
\begin{eqnarray}
 d &=&  a \approx c\left(\frac{dr^2-p_b^2}{dr}-\frac{p_b^2}{dr(dr^2-p_b^2)}p_l^2+{\mathcal O}(p_l^4)\right),\nonumber\\
   b&=&c p_l^2.
\end{eqnarray}
Finally, the transformed root is
\begin{equation}
  p_s=\frac{p_b^2-dr^2}{p_b}+\frac{p_b^2}{p_b(p_b^2-dr^2)}p_l^2+{\mathcal O}(p_l^4).
\end{equation}
The asymptotic expression for the complete elliptic integral when its 
argument is small is 
\beq
  {\cal K}(k') 
  \sim \frac{1}{2}\ln\left(\frac{16}{k'^2}\right). \label{eqn:el_ass}
\enq

Wrapping up all of the above, we obtain the full time evolution 
$p(t)$ from Eq.~(\ref{eq:full_solution}) where we substitute
\begin{eqnarray}
  A^2&=&1-(\frac{dE}{K})^2,\nonumber\\
  p_s&=& \frac{2K}{I_0} (1-(\frac{dE}{K})^2),
\end{eqnarray}
and $p_0$ is the initial population; the time taken to attain the 
first maximum is given by
\beq
  T = \frac{1}{4K\sqrt{1-(\frac{dE}{K})^2}}\ln\left(\frac{16}{k'^2}\right),
\label{maxtime}
\enq
where
\beq
  k'^2\approx\frac{1}{(1-(\frac{dE}{K})^2)^2}\frac{I_0^2}{
  4(K+dE)^2}p_0^2.
\enq

  %
%
\end{document}